\documentclass[journal,twocolumn,final]{IEEEtran}

\usepackage[english]{babel}
\usepackage{cuted}
\usepackage{flushend}
\usepackage{amsthm}
\usepackage{amsmath}
\usepackage{amsfonts}
\usepackage{dsfont}
\usepackage{mathrsfs}
\usepackage{pifont}
\usepackage{amssymb}
\usepackage{verbatim}
\usepackage{upgreek}
\usepackage{color}
\usepackage[final]{graphicx}
\usepackage{caption}
\usepackage{algorithmic}
\usepackage{algorithm}
\usepackage{epsfig}
\usepackage{eucal}
\usepackage{cite}
\usepackage{subfigure}
\usepackage{stmaryrd}

\makeatletter
\def\maketag@@@#1{\hbox{\m@th\normalfont\normalsize#1}}
\makeatother

\newcommand{\bbm}{\begin{bmatrix}}
\newcommand{\ebm}{\end{bmatrix}}

\newcommand{\bit}{\begin{itemize}}
\newcommand{\eit}{\end{itemize}}

\newcommand{\ben}{\begin{enumerate}}
\newcommand{\een}{\end{enumerate}}

\newcommand{\bdesc}{\begin{description}}
\newcommand{\edesc}{\end{description}}

\newcommand{\bea}{\begin{array}}
\newcommand{\eea}{\end{array}}

\newcommand{\tr}{\mbox{\rm Tr}\, }

\newcommand{\beqa}{\begin{eqnarray}}
\newcommand{\eeqa}{\end{eqnarray}}

\newcommand{\ds}{\displaystyle}

\newcommand{\Comment}[1]{}

\def\R{{\mathds R}}
\def\Real{{\mathfrak Re}}
\def\C{{\mathds C}}

\def\cA{\mbox{$\mathcal A$}}

\def\ccC{\mbox{$\mathcal C$}}
\def\cC{\mbox{$\CMcal C$}}

\def\cL{\mbox{$\mathcal L$}}
\def\cN{\mbox{$\CMcal N$}}

\newcommand{\barl}{\bar{l}}

\newcommand{\be}{\begin{equation}}
\newcommand{\ee}{\end{equation}}

\newcommand{\bzero}{{\mbox{\boldmath $0$}}}

\newcommand{\bm}{{\mbox{\boldmath $m$}}}
\newcommand{\bp}{\mbox{\boldmath $p$}}

\newcommand{\bx}{{\mbox{\boldmath $x$}}}
\newcommand{\by}{{\mbox{\boldmath $y$}}}

\newcommand{\bz}{{\mbox{\boldmath $z$}}}

\newcommand{\bA}{{\mbox{\boldmath $A$}}}
\newcommand{\bB}{{\mbox{\boldmath $B$}}}
\newcommand{\bC}{{\mbox{\boldmath $C$}}}

\newcommand{\bE}{{\mbox{\boldmath $E$}}}

\newcommand{\bI}{{\mbox{\boldmath $I$}}}
\newcommand{\bJ}{{\mbox{\boldmath $J$}}}

\newcommand{\bM}{{\mbox{\boldmath $M$}}}

\newcommand{\bS}{{\mbox{\boldmath $S$}}}
\newcommand{\bT}{{\mbox{\boldmath $T$}}}
\newcommand{\bU}{{\mbox{\boldmath $U$}}}
\newcommand{\bV}{{\mbox{\boldmath $V$}}}

\newcommand{\bZ}{{\mbox{\boldmath $Z$}}}

\DeclareMathOperator*{\argmax}{arg\,max}

\newcommand{\btheta}{{\mbox{\boldmath $\theta$}}}

\newcommand{\dmax}{\begin{displaystyle}\max\end{displaystyle}}

\newcommand{\testest}{\mbox{$
		\begin{array}{c}
		\stackrel{ \stackrel{\textstyle H_{1,\widehat{m}}}{\textstyle >} }{
			\stackrel{\textstyle <}{\textstyle H_0} }
		\end{array}
		$}}

\begin{document}

\title{EM-based Solutions for Covariance Structure Detection and Classification in Polarimetric SAR Images}

\author{Pia Addabbo, \emph{Senior Member, IEEE}, Filippo Biondi, \emph{Member, IEEE}, 
Carmine Clemente, \emph{Senior Member, IEEE}, Sudan Han, 
Danilo Orlando, \emph{Senior Member, IEEE},  and Giuseppe Ricci, \emph{Senior Member, IEEE}
\thanks{Pia Addabbo is with Universit\`a degli studi ``Giustino Fortunato'', Benevento, Italy. E-mail: {\tt p.addabbo@unifortunato.eu}.}
\thanks{Fillippo Biondi is with Italian Ministry of Defence. Email: {\tt biopippo@gmail.com}.}
\thanks{Carmine Clemente is with the University of Strathclyde, Department of Electronic and Electrical Engineering, 204 George Street, 
G1 1XW, Glasgow, Scotland. E-mail: {\tt carmine.clemente@strath.ac.uk}}
\thanks{S. Han is with the National Innovation Institute of Defense Technology, Beijing, China E-mail: {\tt xiaoxiaosu0626@163.com}.}
\thanks{D. Orlando is with Universit\`a degli Studi ``Niccol\`o Cusano'', 00166 Roma, Italy. E-mail: {\tt danilo.orlando@unicusano.it}.}
\thanks{G. Ricci is with the Dipartimento di Ingegneria dell'Innovazione, Universit\`{a} del Salento, 
Via Monteroni, 73100 Lecce, Italy. E-mail: {\tt giuseppe.ricci@unisalento.it}.}
}

\maketitle

\begin{abstract}
This paper addresses the challenge of classifying  polarimetric SAR images by leveraging the peculiar 
characteristics of the polarimetric covariance matrix (PCM). To this end, a general framework to solve 
a multiple hypothesis test is introduced with the aim to detect and classify contextual spatial variations in polarimetric SAR images. 
Specifically, under the null hypothesis, only an unknown structure is assumed for data belonging to a $2$-dimensional
spatial sliding window, whereas under each alternative hypothesis, data are partitioned into subsets sharing different structures.
The problem of partition estimation is solved by resorting to hidden random variables representative of covariance
structure classes and the expectation-maximization algorithm. The effectiveness of the proposed detection strategies 
is demonstrated on both simulated and real polarimetric SAR data also in comparison with existing classification algorithms. 
\end{abstract}


\section{Introduction}
In the last 20 years, the benefits of information extraction from synthetic aperture radar (SAR) \cite{9019638,9160246,6784019} 
and, in particular, polarimetric SAR images have been widely demonstrated in a range of applications including 
environmental monitoring \cite{6716971, 6562763}, security \cite{570713,6943309} and urban area monitoring \cite{1580721, 5475228}. 
Thanks to the increasing number of use cases for this specific type of sensor, more and more current and future 
remote sensing missions use polarimetric SAR sensors, despite their increased costs. 
A key aspect of polarimetric SAR is the capability to extract information about the scattering mechanisms of the scene of interest, 
thus allowing for a more advanced characterization of the scene. Specifically, the polarimetric scattering phenomenon of 
a medium can be completely described by using the covariance matrix  \cite{symmetriesPol}. Generally speaking, 
symmetric properties arise in the encountered medium, which are, in principle, detectable through the related 
covariance matrix form. However, there exist different (spatially distributed) forms for the structure of the covariance matrix
depending on the nature of the imaged scene. As a matter of fact, when applying polarimetric covariance matrix (PCM) 
based image classification approaches such as the one proposed in \cite{DetectionSymmetries}, it might frequently occur 
that inhomogeneous areas are under analysis. Even though such areas contain a mixture of covariance matrix symmetries, methods of
\cite{DetectionSymmetries} detect only the dominant symmetry. It naturally turns out that more information can be extracted 
if the presence of different symmetries can be identified within the window under test.

With the above remarks in mind, in this paper a contextual approach aimed at detecting the changes in the structure 
of the PCM between neighbor cells under test is proposed. To be more precise, the proposed framework considers 
the same polarimetric covariance structures as in \cite{DetectionSymmetries,robustFramework} and formulates the problem as 
a multiple hypothesis test, where, unlike \cite{DetectionSymmetries,robustFramework}, data under test might not share
the same PCM structure. In fact, as shown in Section \ref{sec:problemFormulation}, the detection problem at hand contains only 
one null hypothesis, where all the cells under test exhibit the same (unknown) PCM structure, and multiple alternative
hypotheses accounting for at least two different (and unknown) PCM structures. The number of alternative hypothesis depends
on the entire set of considered structures for the PCM. In addition, under the generic alternative hypothesis, a data partition
is accomplished in order to identify the subsets of cells with a specific PCM structure. In this respect, notice that the maximum 
likelihood approach (MLA) would lead to very time demanding estimation procedures since for each combination of the 
available PCM structures, a maximization over all the possible partitions should be performed. For this reason, an alternative 
approach, grounded on the equivalence between partitioning and labeling, is pursued. Specifically, the classification task
is accomplished by introducing hidden random variables that are representative of the different PCM structure classes, and
estimating the resulting unknown parameters through the expectation-maximization (EM) algorithm \cite{Dempster77}.
This approach to PCM classification appears here for the first time (at least to the best of authors' knowledge) and represents 
the main technical novelty of this paper. Finally, the decision statistic is built up by leveraging the innovative design framework developed
in \cite{penalizedLLRT_part1} where the log-likelihood ratio test (LLRT) is adjusted by means of suitable penalty terms
borrowed from the model order selection (MOS) rules \cite{Stoica1}.

The remainder of this paper is organized as follows. The next section formally introduces the multiple hypothesis test defining
the measurement models as well as the unknown parameters. Section \ref{sec:designs} describes the estimation procedures
along with the design of the detection architectures. Illustrative examples based upon both simulated and real-recorded data
are confined to Section \ref{sec:numericalExamples}, whereas concluding remarks and possible future research 
lines are contained in Section \ref{sec:conclusions}.

\subsection{Notation}
In the sequel, vectors and matrices are denoted by boldface lower-case and upper-case letters, respectively.
The symbols $\det(\cdot)$, $\tr(\cdot)$, $(\cdot)^T$, and $(\cdot)^\dag$ denote the determinant, 
trace, transpose, and conjugate transpose, respectively. As to the numerical sets, $\R$ is the set of 
real numbers, $\R^{N\times M}$ 
is the Euclidean space of $(N\times M)$-dimensional real matrices (or vectors if $M=1$), $\C$ is the set of 
complex numbers, and $\C^{N\times M}$ is the Euclidean space of $(N\times M)$-dimensional complex matrices 
(or vectors if $M=1$). If $A$ and $B$ are two sets, $A\setminus B$ is the set containing the elements of $A$
that do not belong to $B$; the empty set is denoted by $\emptyset$.
%The cardinality of a set $\Omega$ is denoted by $|\Omega|$.
The modulus of $x\in\C$ is denoted by $|x|$, whereas symbol $\propto$ means proportional to.
%The Dirac delta function is indicated by $\delta(\cdot)$.
Symbol $\Real\left\{ z \right\}$ indicates the real part of the complex number $z$.
The acronyms PDF and IID mean probability density function and independent and identically distributed, respectively.
$\bI$ and $\bzero$ stand for the identity matrix and the null vector/matrix of proper size, respectively. 
Finally, we write $\bx\sim\cC\cN_N(\bm, \bM)$ if $\bx$ is a complex circular $N$-dimensional normal 
vector with mean $\bm$ and positive definite covariance matrix $\bM$.

%%%%%% while $\bx\sim\cC\cA\cG_N(\bzero,\bM)$ 
%%%%%%means that $\bx\in\C^{N\times 1}$ obeys the central complex angular Gaussian distribution 
%%%%%%with $\bM$ a positive definite matrix.

\section{Sensor Model and Problem Formulation}
\label{sec:problemFormulation}

\begin{figure}[tbp!]
    \centering
    \includegraphics[width=.45\textwidth]{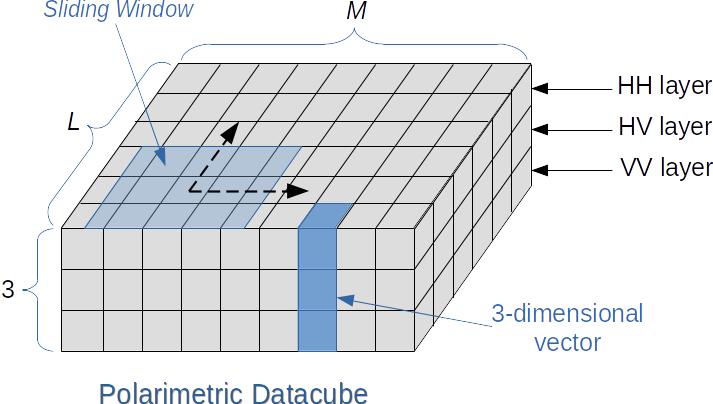}
    \caption{Schematic representation of a polarimetric SAR image as a datacube and sliding window used to obtain data under test.}
    \label{fig:polDataCube}
\end{figure}

A multipolarization SAR sensor generates an
image (datacube) where each pixel is represented by a vector whose 
entries are the complex returns corresponding to the 
different polarimetric channels. Here, 
we assume that the medium is reciprocal allowing to deal with the three polarimetric channels
HH, HV, and VV \cite{symmetriesPol}.
Let us denote by $L$ and $M$ the numbers of pixels along the vertical and horizontal dimensions of the polarimetric image, respectively, 
then, the sensor provides a datacube of size $L\times M\times 3$ (see Figure \ref{fig:polDataCube}). 
Now, the set of vectors under test is selected using a sliding window that moves over the image and contains 
$K$ statistically independent random vectors $\bz_k\in\C^{3\times 1}$, $k=1,\ldots,K$, such
that $\bz_k\sim\cC\cN_3(\bzero, \bM_k)$, with $\bM_k\in\C^{3\times 3}$, $k=1,\ldots,K$, the positive definite 
unknown PCM. Moreover, the latter
exhibits specific configurations according to the scattering mechanisms in play \cite{lee2017polarimetric,symmetriesPol}. 
Specifically, given the $k$th vector, the polarimetric structure takes on the following forms:
\begin{itemize}
\item in the presence of a reciprocal medium, we have that
\be\label{eqn:polCov_1}
{\bM}_k=\begin{bmatrix}
	c_{hhhh} & c_{hhhv} & c_{hhvv} \\
	c^{*}_{hhhv} & c_{hvhv} & c_{hvvv} \\
	c^{*}_{hhvv} & c^{*}_{hvvv} & c_{vvvv} 
\end{bmatrix}=\bC_1;
\ee
\item in the presence of a reflection symmetry with respect to a vertical plane, the structure becomes
\be\label{eqn:polCov_2}
{\bM}_k=\begin{bmatrix}
	c_{hhhh} & 0 & c_{hhvv} \\
	0 & c_{hvhv} & 0 \\
	c_{hhvv}^* & 0 & c_{vvvv} 
\end{bmatrix}=\bC_2;
\ee
\item when a rotation symmetry is present, we can write
\be\label{eqn:polCov_3}
{\bM}_k=\begin{bmatrix}
	c_{hhhh} & c_{hhhv} & c_{hhvv} \\
	-c_{hhhv} & c_{hvhv} & c_{hhhv} \\
	c_{hhvv} & -c_{hhhv} & c_{hhhh} 
\end{bmatrix}=\bC_3,
\ee
where $\Real\{c_{hhhv}\}=0$, $c_{hhvv}\in \R$ and $c_{hvhv}=(c_{hhhh}-c_{hhvv})/2$;
\item in the case of azimuth symmetry, it is given by
\be\label{eqn:polCov_4}
{\bM}_k=\begin{bmatrix}
	c_{hhhh} & 0 & c_{hhvv} \\
	0 & c_{hvhv} & 0 \\
	c_{hhvv} & 0 & c_{hhhh} 
\end{bmatrix}=\bC_4,
\ee
where $c_{hhvv}\in\R$ and $c_{hvhv}=(c_{hhhh}-c_{hhvv})/2$.
\end{itemize}
It turns out that within the sliding window containing the vectors under test, 
several situations may occur according to the involved structures. 
Specifically, the PCM can remain unaltered within the sliding window 
or at least two different forms appear within the window.

To be more precise, we are interested in distinguishing the 
case $\bM_1=\ldots=\bM_K\in\ccC=\{\bC_1,\ldots,\bC_4\}$ from different configurations  where the pixels are characterized by at least two PCMs.
This problem can be formulated in terms of a multiple hypothesis test consisting
of one null hypothesis and several alternative hypotheses, namely as
%%%%%. Starting from the H-RE, 
%%%%%let us notice that the Probability Density Function (PDF) of $\bn_{1}\sim\cC\cN_3(\bzero,s_1\bSigma_1)$ is the same as 
%%%%%that of $\bn_2\sim\cC\cN_3(\bzero,s_2\bSigma_2)$, where $s_2=s_1/A$ and $\bSigma_2=A\bSigma_1$ with $A>0$ an arbitrary constant.
%%%%%It follows that the parameters $(s_k,\bSigma_k)$ are not identifiable. 
%%%%%Otherwise stated, these parameters can be estimated up to arbitrary constants under
%%%%%each hypotheses. This ambiguity can be removed by setting $s_k=1$, $k=1,\ldots,K$, yielding $\bM_k=\bSigma_k$, and the problem at 
%%%%%hand can be written as
\be\!\!\!
\begin{cases}
	H_0 : &\!\!\!\!\!\! \ds\bz_k \sim \cC\cN_3(\bzero, \bC_{i_0}), i_0 \in \{1, \ldots, 4\},
	\\
	H_{{1,1}} : & \!\!\!\!\!\! \begin{cases}
		\bz_k \sim \cC\cN_3(\bzero, \bC_{i_0}), k \in \Omega_{1}\subset \Omega, \\
		\bz_k \sim \cC\cN_3(\bzero, \bC_{i_1}), k \in \Omega_{2}=\Omega \setminus \Omega_{1}, \\
		i_0 < i_1, i_0,i_1 \in \{1, \ldots, 4\},
	\end{cases}	
	\\
	H_{{1,2}} : & \!\!\!\!\!\! \begin{cases}
		\bz_k \sim \cC\cN_3(\bzero, \bC_{i_0}), k \in \Omega_{1}\subset \Omega, \\
		\bz_k \sim \cC\cN_3(\bzero, \bC_{i_1}), k \in \Omega_{2}\subset \Omega\setminus\Omega_{1}, \\
		\bz_k \sim \cC\cN_3(\bzero, \bC_{i_2}), k \in \Omega_{3}= \Omega\setminus \{\Omega_{1} \cup \Omega_{2}\}, \\
		i_0 < i_1 < i_2, i_0,i_1,i_2 \in \{1, \ldots, 4\},
	\end{cases}
	\\
	H_{{1,3}} : & \!\!\!\!\!\! \begin{cases}
		\bz_k \sim \cC\cN_3(\bzero, \bC_{1}), k \in \Omega_{1}\subset \Omega, \\
		\bz_k \sim \cC\cN_3(\bzero, \bC_{2}), k \in \Omega_{2}\subset \Omega\setminus\Omega_{1}, \\
		\bz_k \sim \cC\cN_3(\bzero, \bC_{3}), k \in \Omega_{3}\subset \Omega\setminus \{\Omega_{1} \cup \Omega_{2}\}, \\
		\bz_k \sim \cC\cN_3(\bzero, \bC_{4}), k \in \Omega_{4}= \Omega\setminus \{\Omega_{1} \cup \Omega_{2} \cup \Omega_3 \}, 
		%\\
		%\bM_1\neq\bM_2\neq\bM_3\neq\bM_4,
	\end{cases}
\end{cases}
\label{eqn:MHP1}
\ee
where $\Omega=\{1,\ldots,K\}$ and
the $\Omega_l$s are unknown (except for $\Omega_{i+1}$ under $H_{1,i}$).
The PDF of $\bZ=[\bz_1,\ldots,\bz_K]$ under
$H_0$ is given by
\be
p_0(\bZ;\bC_{i_0}) = \frac{\exp\left\{-\tr\left[\bC_{i_0}^{-1}\bZ\bZ^\dag\right]\right\}}{\pi^{3K} \left[\det\left (\bC_{i_0} \right)\right]^{K}},
\label{eqn:pdf_H0_hom}
\ee
whereas that under $H_{1,m}$, $m=1,\ldots,3$, can be written as
\begin{multline}
p_{1,m}(\bZ;\bC_{i_0},\ldots,\bC_{i_{m}}) 
\\
=\prod_{l=0}^{m}\prod_{k\in\Omega_{l+1}}
\frac{\exp\left\{-\tr\left[\bC_{i_l}^{-1}\bz_k\bz_k^\dag\right]\right\}}{\pi^{3} \left[\det(\bC_{i_l})\right]}
\label{eqn:pdf_H1i_hom}
\end{multline}
with the constraints 
\be
\bigcup_{l=0}^{m}\Omega_{l+1}=\Omega \quad \mbox{and} \quad  \Omega_i\cap\Omega_j=\emptyset,  \ i\neq j.
\ee
For future reference, it is also useful to define
the sets
\be
\cA_m=\{i_0,\ldots,i_{m}\}\subseteq \{1,\ldots,4\}, \ m=0,\ldots, 3,
\ee
and\footnote{Notice that $\cA_3=\{1,2,3,4\}$.}
denote by $\btheta_0(\cA_0)$ and $\btheta_1(\cA_m)$ the unknown parameters under $H_0$, given $\cA_0$, and under $H_{1,m}$, given $\cA_m$,
respectively.

\section{Detection Architecture Designs}
\label{sec:designs}
In this section, we provide some important remarks that are preparatory to the subsequent derivations and motivate the design choices.
As specified below, the adopted decision rules rely on the LLRT where the unknown parameters are replaced by suitable estimates. 
However, implementation of such a strategy for the problem at hand requires to circumvent two main drawbacks.

First of all, under $H_{1,m}$, 
the partition $\left\{\Omega_1,\ldots, \Omega_{m+1}\right\}$ of the pixels of the sliding window is not known. 
As a consequence, application of the MLA
to obtain the parameter estimates would be a formidable task:
we should consider all the combinations of $m+1$ PCM structures over the 
available options, namely ${4}\choose{m+1}$, and for each of them the different partitions of $\Omega$ into $m+1$ subsets.
Therefore, in what follows, we propose
two alternative solutions that abstain from the computation of all the possible partitions of $\Omega$.
These alternatives rely on the fact that, from an operating
point of view, partitioning $\Omega$ is tantamount to labeling its elements.  
Therefore, we can follow the lead of \cite{9321174} and introduce
$K$ IID hidden discrete random variables that are representative of the 
labels associated with the $\bz_k$s under $H_{1,m}$
and $\cA_m=\{i_0,\ldots,i_{m}\}\subseteq \{1\ldots,4\}$. In fact, such random variables take on values in
$\cA_m$.
Then, we apply the EM algorithm \cite{Dempster77} to estimate the unknown parameters.
The herein proposed estimation procedures differ from each other in the way such hidden random variables
are defined and used to build up the LLRT, a point better specified at the end of this section.

The second drawback of implementing a plain LLRT is originated by the fact that 
the elements of
$\ccC$ are characterized
by different numbers of unknowns. Thus, not only a balanced comparison of the hypotheses, but also of 
the different $\cA_m$s, given $m$ (i.e., given the hypothesis), requires introducing 
adequate penalty factors. To be more quantitative, we observe that
the number $n_i$ of unknown parameters associated with $\bC_i$, $i=1,\ldots,4$, is given by
\be
n_i=\begin{cases}
9 & \mbox{if } i=1,
\\
5 & \mbox{if } i=2,
\\
3 & \mbox{if } i=3,
\\
2 & \mbox{if } i=4.
\end{cases}
\ee
Accordingly, the number of unknowns associated with
$\cA_m=\{i_0,\ldots,i_{m}\}$ can be computed as
$u\left(\cA_m\right)=\sum_{j=0}^{m} n_{i_j}$.

With the above remarks in mind, we devise decision schemes for problem \eqref{eqn:MHP1} exploiting a penalized 
LLRT \cite{van2004optimum}. As a first step towards the introduction of such a  penalized LLRT, we 
denote by $\widehat{\btheta}_0(\cA_0)$ ($\widehat{\btheta}_1(\cA_m)$) the estimate of the unknown parameters related to $H_0$ and $\cA_0$ ($H_{1,m}$ and $\cA_m$). 
Similarly, $\widehat{\btheta}_{1}(\cA_{\widehat{m}})$
is the estimate of the unknown parameters associated with $H_{1,\widehat{m}}$ and $\cA_{\widehat{m}}$.
For the moment we leave aside the description of the  estimation procedures, which will be the object of the next subsections, and
introduce the general structure of the penalized LLRT
\begin{align}
\nonumber
&\max_{m \in \{ 1,2,3\}} \max_{\cA_m}
\left[
\log g_{1}\left(\bZ;\widehat{\btheta}_{1}\left(\cA_m\right)\right)
-h_1\left(\cA_m\right)
\right]
\\ 
&-
\max_{\cA_0}
\left[
\log g_{0}\left(\bZ;\widehat{\btheta}_{0}\left(\cA_0\right)\right)
-h_0\left(\cA_0\right)
\right]
\testest\eta,
\label{eqn:penalizedLLRT}
\end{align}
where 
\[
\widehat{m}=\argmax_{m \in \{ 1,2,3\}}
\left\{\max_{\cA_m}
\left[
\log g_{1}\left(\bZ;\widehat{\btheta}_{1}\left(\cA_m\right)\right)
-h_1\left(\cA_m\right)
\right]\right\},
\]
$g_{0}\left(\bZ;\btheta_{0}\left(\cA_0\right)\right)
=p_0(\bZ;\bC_{i_0})$, $g_{1}\left(\bZ;\btheta_{1}\left(\cA_m\right)\right)$
%=p_{1,m}(\bZ;\bC_{i_0},\ldots,\bC_{i_{m}}),
denotes the PDF of the observables under $H_{1,m}$ and $\cA_m$, that will be specified by subsequent sections,
$h_1\left(\cA_m\right)$, $m=1,2,3$, is a penalty term  accounting for the number of unknown parameters
related to $H_{1,m}$ and $\cA_m$, $h_0\left(\cA_0\right)$
is a penalty term  accounting for the number of unknown parameters
related to $H_{0}$ and $\cA_0$,
and $\eta$ is the detection 
threshold\footnote{Hereafter, we denote by $\eta$ the generic detection threshold.} to be set according to 
the probability of false alarm ($P_{fa}$). The penalty terms can be written as $h_1(\cA_m)=\gamma \left( u\left(\cA_m\right) +m+1 \right)$
and $h_0(\cA_0)=\gamma u\left(\cA_0\right)$
 where we recall that $u\left(\cA_m\right)$ 
is the number of unknown real-valued parameters associated with $\cA_m$,
$m+1$ is the number of unknowns related to the 
probability mass function (PMF)
of
the hidden discrete random variables (such random variables take on values in
$\cA_m$),
and $\gamma$ is a factor borrowed from the MOS rules \cite{Stoica1} 
as the Akaike Information Criterion (AIC), the Generalized Information Criterion (GIC), and the Bayesian Information Criterion (BIC), i.e.,
\be
\gamma=
\begin{cases}
1, & \mbox{for AIC-based Detector (AIC-D)},
\\
\log(6K)/2, & \mbox{for BIC-based Detector (BIC-D)},
\\
(1+\rho)/2, \ \rho>1, & \mbox{for GIC-based Detector (GIC-D)}.
\end{cases}
\label{eqn:penaltyTerm}
\ee
It is important to stress that, under $H_{1,m}$ and 
$\cA_m$, $u\left(\cA_m\right)$
is obtained by partitioning the data set into $m+1$ subsets, associating
with them specific structures, and summing the respective number of unknown parameters. 
The cardinality of each subset
along with the coordinates of the vectors within it are also unknowns, but they are independent of $\cA_m$ and, hence,
irrelevant to the decision process. 
%because if we assign to each vector 
%a label that classifies it as belonging to a specific subset, then the partitioning problem 
%is tantamount to estimating the labels associated with each vector.
%As a consequence, the total number of labels is equal to the number of vectors $K$ and does not depend on $H_{1,m}$.

It still remain to show how to estimate
$\btheta_0(\cA_0)$ and $\btheta_1(\cA_m)$.
%It turns out that such strategy is unacceptable due to the huge computational burden. For this reason, in what follows, we propose
%two alternative solutions that abstain from the computation of all the possible partitions of $\Omega$.
%Specifically, they rely on the previous remark that partitioning $\Omega$ from an operating
%point of view is tantamount to labeling its elements.  
%Therefore, 
As previously anticipated, we will follow the lead of \cite{9321174} and introduce
$K$ independent and identically distributed hidden discrete random variables that ``specify the characterization'' of the
$\bz_k$s. Then, we apply the EM algorithm \cite{Dempster77} to estimate the unknown parameters.
The herein proposed estimation procedures differ from each other in the way such hidden random variables
are defined and used to build up the LLRT under $H_{1,m}$.

The first procedure assumes that under $H_{1,m}$ the hidden random variables, $c_{k,m}$ say, have alphabet 
$\cA_m=\{i_0,\ldots,i_{m}\}\subseteq \{1\ldots,4\}$ with PMF
\be
\begin{cases}
P(c_{k,m}=l)=P_{l,m}, \ l\in\cA_m,
\\
\ds\sum_{l\in{\mathcal A}_m} P_{l,m}=1,
\end{cases}
\ee
and that when $c_{k,m}=l$, $l\in\cA_{m}$, 
then $\bM_k=\bC_l$. Therefore, we can write the PDF of $\bz_k$ under $H_{1,m}$ as \cite{9321174}
\be
f_{1,m}(\bz_k;\btheta_1(\cA_m))=\sum_{l\in {\mathcal A}_m} P_{l,m}f(\bz_k; \bC_l),
\label{eqn:LVM_PDF}
\ee
where $f(\bz_k;\bC_l)$ is the PDF of $\bz_k\sim\cC\cN(\bzero,\bC_l)$.
%and, with a little abuse of notation, 
%$\btheta_1(\cA_m)$ denotes the unknown parameter vector %under $H_{1,m}$.\footnote{Hereafter, we use symbol
%$\btheta_1(\cA_m)$ to indicate the generic parameter %vector.}
The above PDF will be used in place of the original PDF to form the LLRT. Notice that 
$\btheta_1(\cA_m)$ depends on the specific choice for the alphabet of the hidden random variables.
As a matter of fact, for each alternative hypothesis, each of the ${4}\choose{m+1}$ combinations of the PCM structures 
identifies an alphabet configuration. Thus, we come up with $6$, $4$, and $1$ different alphabet configurations 
under $H_{1,1}$, $H_{1,2}$, and $H_{1,3}$, respectively. Nevertheless, as we will show in the next subsections,
these configurations can be handled without a dramatic increase of the computational requirements.

The second approach does not account for the hypotheses $H_{1,1}$, $H_{1,2}$, and $H_{1,3}$ to set 
the number of classes but simply considers all classes. As consequence,
the hidden random variables, $c_k$ say, share the same alphabet $\cA=\{1,2,3,4\}$ and PMF $P(c_{k}=l)=P_{l}$, $l\in\cA$.
The LLRT under $H_{1,m}$ is formed by selecting the $m+1$ 
most probable PCM structures and modifying \eqref{eqn:LVM_PDF} according to
the selected structures.

In the next subsections, we describe in the detail these procedures that are based upon the EM algorithm.

\subsection{First EM-based Estimation Strategy}
Let us assume that 
under $H_{1,m}$, $m=1,2,3$, equation \eqref{eqn:LVM_PDF} holds true and focus on problem \eqref{eqn:MHP1}.
Now, given a configuration for $\cA_m$, the 
log-likelihood of $\bZ$ is given by
\be
\log g_{1}\left(\bZ;\btheta_{1}\left(\cA_m\right)\right)
%\cL_m(\btheta_1(\cA_m);\bZ)
=\sum_{k=1}^K\log \sum_{l\in{\mathcal A}_m} P_{l,m}f(\bz_k; \bC_l).
\label{eqn:jointLL}
\ee
The application of the EM algorithm consists of the E-step that leads to \cite{9321174,murphy2012machine}
\be
q_k^{(h-1)}(l,m)=\frac{f(\bz_k;\widehat{\bC}_{l}^{(h-1)})\widehat{P}_{l,m}^{(h-1)}}
{\ds\sum_{n\in{\mathcal A}_m}f(\bz_k;\widehat{\bC}_{n}^{(h-1)})\widehat{P}_{n,m}^{(h-1)}}, \quad l\in\cA_m, 
\label{eqn:E-step_1_H-RE}
\ee
where $\widehat{P}_{l,m}^{(h-1)}$, $l\in\cA_m$, and $\widehat{\bC}_{n}^{(h-1)}$, $n\in\cA_m$, are the available estimates 
at the $(h-1)$th step,
and of the M-step requiring to solve the following joint optimization problem\footnote{For brevity, we have 
omitted some derivation details of the EM algorithm and refer the interested reader 
to \cite{9321174,murphy2012machine} for further information.}
\begin{multline}
\dmax_{{\bf p}_m} \dmax_{{\bf C}_l \atop l\in {\mathcal A}_m} 
\bigg\{ \sum_{k=1}^{K} \sum_{l\in{\mathcal A}_m} q_k^{(h-1)}(l,m) [ -\log \det(\bC_{l})
\\ 
-\tr(\bC_{l}^{-1} \bz_k \bz_k^\dag)]  -  \sum_{k=1}^{K} \sum_{l\in{\mathcal A}_m} q_k^{(h-1)}(l,m)  \log P_{l,m}\bigg\},
\label{eqn:M-step_1_H-RE}
\end{multline}
where\footnote{Notice that the entries of $\bp_m$ are nonnegative.} $\bp_m=[P_{i_0,m},\ldots,P_{i_{m},m}]^T\in\R^{(m+1)\times 1}$.

It is not difficult to show that the maximization with respect to $\bp_m$, accomplished under the constraint 
\be
\sum_{l\in\cA_m} P_{l,m}=1,
\ee
returns the following stationary points
\be
\widehat{P}_{l,m}^{(h)}=\frac{1}{K}\sum_{k=1}^{K}q_k^{(h-1)}(l,m), \quad l\in\cA_m.
\label{eqn:priorsEstimates}
\ee
On the other hand, the maximization with respect to $\bC_l$ for a given $\barl\in\cA_m$ implies
\be
\dmax_{\bC_{\barl}} \sum\limits_{k=1}^{K} q_k^{(h-1)}(\barl,m)\left[ 
-\log \det(\bC_{\barl}) -\tr\left( \bC_{\barl}^{-1} \bz_{k}\bz_{k}^{\dag} \right) \right].
\label{eqn:optmizationGeneral}
\ee
Let us solve the above problem for each possible value taken on by $\barl$. To this end, notice that
when $\barl=1$, we have that
\begin{align}
&\sum_{k=1}^{K} q_k^{(h-1)}(1,m)\left[ 
-\log \det(\bC_{1}) -\tr\left( \bC_{1}^{-1} \bz_{k}\bz_{k}^{\dag} \right) \right]\nonumber
\\
&=q^{(h-1)}(1,m)\left\{\log \det(\bC_{1}^{-1})-\tr\left[ \bC_{1}^{-1}\bS_q(1,m)^{(h-1)} \right]\right\},
\end{align}
where $q^{(h-1)}(1,m)=\sum_{k=1}^{K} q_k^{(h-1)}(1,m)$, $\bS_q(1,m)^{(h-1)}=\sum_{k=1}^{K} q_k^{(h-1)}(1,m)\bz_k\bz_k^\dag /q^{(h-1)}(1,m)$.
It follows that
\be
\dmax_{\bC_1} \log \det(\bC_{1}^{-1})-\tr\left[ \bC_{1}^{-1}\bS_q(1,m)^{(h-1)} \right]
\ee
is tantamount to maximize
\begin{multline}
\log \det\left(\bC_{1}^{-1}\bS_q(1,m)^{(h-1)}\right)-\log\det\left(\bS_q(1,m)^{(h-1)}\right)
\\
-\tr\left[ \bC_{1}^{-1}\bS_q(1,m)^{(h-1)} \right].
\end{multline}
The maximizer can be obtained resorting to the following inequality \cite{lutkepohl1996handbook}
\be
\log\det(\bA)\leq \tr[\bA]-3,
\label{eqn:inequalityDetTr}
\ee
where $\bA$ is any matrix with nonnegative eigenvalues, and, hence, it follows that
\begin{align*}
\widehat{\bC}_1^{(h)} &=\argmax_{\bC_1} \sum\limits_{k=1}^{K} q_k^{(h-1)}(1,m)
\\
&\times \left[ -\log \det(\bC_{1}) -\tr\left( \bC_{1}^{-1} \bz_{k}\bz_{k}^{\dag} \right) \right]
\\
&=\frac{ \ds\sum_{k=1}^{K} q_k^{(h-1)}(1,m) \bz_k\bz_k^\dagger}{\ds\sum_{k=1}^{K} q_k^{(h-1)}(1,m)}.
\end{align*}

Now, assume that  $\bar{l}=2$ and let $\bU$ be the unitary matrix defined in {\em Lemma 3.1} of \cite{DetectionSymmetries},
then 
\be
\bU\bC_2\bU^\dagger=\begin{bmatrix}
\bA & \bzero\\
\bzero & d
\end{bmatrix},
\ee
where $\bA\in\C^{2\times 2}$ is positive definite and $d>0$. It follows that problem \eqref{eqn:optmizationGeneral} can be recast as
\begin{multline}
\dmax_{\bA}\dmax_{d>0} \sum_{k=1}^{K} q_k^{(h-1)}(2,m) \left[ -\log\det(\bA) -\log d \right. 
\\
\left. -\bz^\dag_{k,1} \bA^{-1} \bz_{k,1} - |z_{k,2}|^2 d^{-1}\right]
\label{eqn:optmizationC2}
\end{multline}
where $\bU \bz_k = [\bz_{k,1}^T \ z_{k,2}]^T$ with $\bz_{k,1}\in \C^{2\times 1}$ and $z_{k,2}\in \C$.
Now, observe that
\be
\lim_{d\rightarrow 0 \atop d\rightarrow +\infty}\left[ -\sum_{k=1}^{K} q_k^{(h-1)}(2,m) (\log d +|z_{k,2}|^2 d^{-1})\right]=-\infty.
\ee
Thus, the stationary points over $d>0$ can be found by setting to zero the first derivative with respect to $d$ of the argument of \eqref{eqn:optmizationC2}, to obtain
\be
-\sum_{k=1}^{K}\frac{{q}_k^{(h-1)}(2,m)}{d}+\frac{1}{d^2}\sum_{k=1}^{K}{q}_k^{(h-1)}(2,m)|z_{k,2}|^2=0.
\ee
Thus, the update of the estimate of $d$ is
\be
\widehat{d}^{(h)}=\frac{\sum\limits_{k=1}^{K}{q}_k^{(h-1)}(2,m)|{z}_{k,2}|^2}{\sum\limits_{k=1}^{K}{q}_k^{(h-1)}(2,m)}.
\ee
As for $\bA$, let us consider
\be
\dmax_{\bA} \sum_{k=1}^{K} q_k^{(h-1)}(2,m) \left[ -\log\det(\bA) -\bz^\dag_{k,1} \bA^{-1} \bz_{k,1}\right],
\label{eqn:optmizationA}
\ee
which can be recast as
\begin{align}
&\dmax_{\bA} q^{(h-1)}(2,m) \log\det(\bA^{-1}) 
- \tr\left[ \bA^{-1} \bS(2,m)^{(h-1)}\right],\nonumber
\\
&\Rightarrow \dmax_{\bA}  \log\det\left[\bA^{-1}{\bS(2,m)^{(h-1)}}/{q^{(h-1)}(2,m)}\right] \nonumber
\\
& \quad -\tr\left[ \bA^{-1}  {\bS(2,m)^{(h-1)}}/{q^{(h-1)}(2,m)}\right],
\label{eqn:optmizationA_1}
\end{align}
where $q^{(h-1)}(2,m)=\sum_{k=1}^{K} q_k^{(h-1)}(2,m)$ and $\bS(2,m)^{(h-1)}=\sum_{k=1}^{K} q_k^{(h-1)}(2,m)\bz_{k,1}  \bz_{k,1}^\dag$.
Exploiting \eqref{eqn:inequalityDetTr}, 
the resulting maximizer for the last problem can be written as
\be
\widehat{\bA}^{(h)}=\frac{\ds\sum_{k=1}^{K} {q}_k^{(h-1)}(2,m)\bz_{k,1}\bz_{k,1}^{\dagger}}
{\ds \sum_{k=1}^{K}{q}_k^{(h-1)}(2,m)}.
\ee
As a consequence, an estimate of $\bC_2$ is given by
\be
\widehat{\bC}_2^{(h)} = \bU^\dag\begin{bmatrix}
	\widehat{\bA}^{(h)} & \bzero\\
	\bzero & \widehat{d}^{(h)}
\end{bmatrix}\bU.
\ee
The next case is $\barl=3$. Notice that matrix $\bC_3$ can be suitably manipulated by applying the transformations
represented by matrices $\bE$, $\bT$, and $\bV$ defined in {\em Lemma 3.1} of \cite{DetectionSymmetries}, namely
\be
\bV \bE \bT \bC_3 \bT^{\dag} \bE \bV^{\dag}= 
\begin{bmatrix}
a & \bzero\\
\bzero & \bB
\end{bmatrix},
\ee
where $a>0$ and $\bB\in\R^{2\times 2}$ is centrosymmetric.\footnote{$\bB$ is such that $\bB=\bJ \bB \bJ$, where
\[
\bJ=\begin{bmatrix}
0 & 1
\\
1 & 0
\end{bmatrix}.
\]}
The objective function can be accordingly expressed as follows
\begin{multline}
\dmax_{a}\dmax_{\bB} \sum_{k=1}^{K} q_k^{(h-1)}(3,m) \left[ -\log a -\log \det(\bB) \right.
\\
\left. - \frac{|x_{k,1}|^2}{a} - \bx_{k,2}^\dag \bB^{-1} \bx_{k,2}  \right],
\label{eqn:case3max}
\end{multline}
where $ \bV \bE \bT \bz_k = [x_{k,1} \ \bx_{k,2}^T]^T$ with $x_{k,1}\in \C$ and $\bx_{k,2}\in \C^{2\times1}$.
Now, since $\bB$ is centrosymmetric, the equality $\bB^{-1}=\left(\bB^{-1} + \bJ \bB^{-1}\bJ \right)/{2}$ holds
and \eqref{eqn:case3max} can be written as 
\begin{multline}
\max_{a}\max_{\bB} \sum\limits_{k=1}^{K} q_k^{(h-1)}(3,m) \left[ -\log a -\log \det(\bB) \right. 
\\
\left. - \frac{|x_{k,1}|^2}{a} -\frac{1}{2} \tr\left[ \bB^{-1} \left( \bx_{k,2} \bx_{k,2}^{\dag} 
+ \bJ \bx_{k,2}\bx_{k,2}^{\dag} \bJ \right) \right] \right].
\end{multline}
Following the same line of reasoning as for the estimation of $d$ and $\bA$, it is possible to show that the estimate of $a$ is
\be
\widehat{a}^{(h)}=\frac{\sum\limits_{k=1}^{K}{q}_k^{(h-1)}(3,m)|x_{k,1}|^2}{\ds\sum_{k=1}^{K}{q}_k^{(h-1)}(3,m)},
\ee
whereas, using \eqref{eqn:inequalityDetTr}, the estimate of $\bB$ has the following expression
\be
\widehat{\bB}^{(h)}=\frac{1}{2}  \frac{\ds \sum_{k=1}^{K}{q}_k^{(h-1)}(3,m)\left(\bx_{k,2}\bx_{k,2}^\dagger+\bJ\bx_{k,2}\bx_{k,2}^\dagger
\bJ\right)}{\ds\sum_{k=1}^{K}{q}_k^{(h-1)}(3,m)}.
\ee
Gathering the above results, we obtain
\be
\widehat{\bC}_3^{(h)} = \bT^\dag\bE^{-1}\bV^{\dag}\begin{bmatrix}
	\widehat{a}^{(h)} & \bzero\\
	\bzero & \widehat{\bB}^{(h)}
\end{bmatrix}\bV\bE^{-1}\bT.
\ee
The final case assumes that $\barl=4$ and $\bC_4$ can be transformed as follows
\be
\mathbf{E}\mathbf{T}\bC_4\mathbf{T}^{\dagger}\mathbf{E}= \begin{bmatrix}
b & 0 & 0 \\
0 & c & 0 \\
0 & 0 & c 
\end{bmatrix} \in \mathbb{R}^{3 \times 3},
\ee
where $b>0$ and $c>0$. As a consequence, the optimization problem to be solved is
\begin{multline}
\dmax_{b>0}\dmax_{c>0} \sum_{k=1}^{K} q_k^{(h-1)}(4,m) \left[ -\log b - 2 \log c \right. 
\\
\left. - \frac{|y_{k,1}|^2}{b} - \frac{1}{c} \by_{k,2}^{\dag} \by_{k,2}  \right]
\end{multline}
where $\bE \bT \bz_k=[ y_{k,1} \ \by_{k,2}^T]^T$ with $y_{k,1}\in \C$ and $\by_{k,2}\in \C^{2\times1}$.
Now, observe that
\begin{align}
\lim_{b\rightarrow 0 \atop b\rightarrow +\infty} \left\{-\sum_{k=1}^{K} q_k^{(h-1)}(4,m) \left[\log b +\frac{|y_{k,1}|^2}{b}\right]\right\}
&=-\infty,
\\
\lim_{c\rightarrow 0 \atop c\rightarrow +\infty} \left\{ -\sum_{k=1}^{K} q_k^{(h-1)}(4,m) \left[2 \log c 
+\frac{1}{c} \by_{k,2}^{\dag} \by_{k,2}\right]\right\} &=-\infty,
\end{align}
and, hence, setting to zero the first derivatives of the above functions with respect to $b$ and $c$, respectively, 
it is not difficult to show that
\be
\widehat{b}^{(h)}=\frac{\sum\limits_{k=1}^{K}{q}_k^{(h-1)}(4,m)|y_{k,1}|^2}{\sum\limits_{k=1}^{K}{q}_k^{(h-1)}(4,m)},
\ee
\be
\widehat{c}^{(h)}=\frac{1}{2}\frac{\sum\limits_{k=1}^{K}{q}_k^{(h-1)}(4,m)\by_{k,2}^\dagger \by_{k,2}}{\sum\limits_{k=1}^{K}{q}_k^{(h-1)}(4,m)}.
\ee
Finally, the estimate of $\bC_4$ is
\be
\widehat{\bC}_4^{(h)} = \bT^\dag\bE^{-1}\begin{bmatrix}
	\widehat{b}^{(h)} & 0 & 0 \\
	0 & \widehat{c}^{(h)} & 0 \\
	0 & 0 & \widehat{c}^{(h)}
\end{bmatrix}\bE^{-1}\bT.
\ee
The actual implementation of the EM algorithm, 
necessary to obtain an estimate of ${\btheta_1}(\cA_m)$,
needs to specify
the convergence criterion that can be used 
to terminate the iterations. In what follows, for each $\cA_m$, $m=1,2,3$, we adopt the following criterion
\begin{multline}
\Delta \cL_m(h)=\Bigg|\Bigg[\cL_m\left(\widehat{\btheta}_1^{(h)}(\cA_m);\bZ\right)
\\
- \cL_m\left(\widehat{\btheta}_1^{(h-1)}(\cA_m);\bZ\right)\Bigg]/{\cL_m\left(\widehat{\btheta}_1^{(h-1)}(\cA_m);\bZ\right)}\Bigg|<\epsilon_m,
\end{multline}
where $\cL_m(\widehat{\btheta}_1^{(h)}(\cA_m);\bZ)=\log g_{1}(\bZ;\widehat{\btheta}^{(h)}_{1}\left(\cA_m\right))$
(see \eqref{eqn:jointLL}) and $\epsilon_m>0$ is set accounting for the requirements in terms of system reactivity. 

The decision statistic of test \eqref{eqn:penalizedLLRT} also requires to estimate the unknown parameters under $H_0$.
The MLE of $\bC_i$ is given by {\em Proposition 3.2} of \cite{DetectionSymmetries}.

\subsection{Second EM-based Estimation Strategy}
The second procedure builds up the term associated with $H_{1,m}$ of the left-hand side of \eqref{eqn:penalizedLLRT} 
by considering the estimates obtained through the first procedure
under $H_{1,3}$ only. Specifically, let us assume that $\cA_3=\{1,2,3,4\}$ and, given $m$, 
select the $m+1$ structures corresponding to the indices
of the $m+1$ highest entries of the final estimate of $\bp_3$ that is denoted 
by $\widehat{\bp}_3=\left[\widehat{P}_{1,3},\widehat{P}_{2,3},\widehat{P}_{3,3},\widehat{P}_{4,3}\right]^T$.

To be more formal, let us sort the $\widehat{P}_{l,3}$s in descending order, namely
\be
\widehat{P}_{l_0,3} \geq \widehat{P}_{l_1,3} \geq \widehat{P}_{l_2,3} \geq \widehat{P}_{l_3,3},
\ee
and form the following subsets $\tilde{\cA}_m=\{l_0,\ldots,l_{m}\}$, $m=1,2,3$, along with the estimate
$\tilde{\btheta}_1(\tilde{\cA}_m)$ that can be drawn from $\widehat{\btheta}_1(\cA_3)$ by picking the components corresponding 
to the indices $l_0,\ldots,l_{m}$. 
Then, decision rule \eqref{eqn:penalizedLLRT} becomes
\begin{align}
\nonumber
&\max_{m \in \{ 1,2,3\}}
\left[
\log g_{1}\left(\bZ;\tilde{\btheta}_{1}\left(\tilde{\cA}_m\right)\right)
-h_1\left(\tilde{\cA}_m\right)
\right]
\\ 
&-
\max_{\cA_0}
\left[
\log g_{0}\left(\bZ;\widehat{\btheta}_{0}\left(\cA_0\right)\right)
-h_0\left(\cA_0\right)
\right]
\testest\eta,
\label{eqn:penalizedLLRT_02}
\end{align}
where
\be
g_{1}\left(\bZ;\tilde{\btheta}_{1}\left(\tilde{\cA}_m\right)\right)=
\sum_{k=1}^K\log \sum_{l\in\tilde{{\mathcal A}}_m} \widehat{P}_{l,m}f(\bz_k; \widehat{\bC}_l).
\ee
The estimation of the unknown parameters under $H_0$ is the same as that of previous subsection.

\subsection{Architecture Summary}
According to the specific penalty term in \eqref{eqn:penalizedLLRT}, that depends on \eqref{eqn:penaltyTerm}, 
and the procedure pursued to come up with the parameter estimates, we can form the following architectures: 
\begin{itemize}
\item AIC-D coupled with procedure $1$ (AIC-D-P1) or procedure $2$ (AIC-D-P2);
\item BIC-D coupled with procedure $1$ (BIC-D-P1) or procedure $2$ (BIC-D-P2); 
\item GIC-D coupled with procedure $1$ (GIC-D-P1) or procedure $2$ (GIC-D-P2).
\end{itemize}
Finally, regardless the estimation procedure applied to data under test, once the unknown parameters have been estimated, data
classification is accomplished as follows
\be
\bz_k \longrightarrow \bC_{\hat{i}} \ \mbox{ if } \hat{i}=\argmax_{l\in\widehat{{\mathcal A}}_{\widehat{m}}}
q_k^{(h_{\max})}(l,\widehat{m}),
\ee
where $h_{\max}$ is the maximum number of EM iterations used to come up with the final parameter estimates, while 
$\widehat{\cA}_m$ and $\widehat{m}$ are the final estimates of $\cA$ and $m$, respectively.

\section{Numerical Examples}
\label{sec:numericalExamples}
In this section, the behaviors of the proposed architectures are assessed using
synthetic data as well as real polarimetric SAR data. Specifically, the first subsection contains the performance results
over simulated data obtained by means of standard Monte Carlo (MC) counting techniques, whereas in Subsection \ref{subsec:realData}, 
the proposed procedures are tested using real polarimetric SAR data. For comparison purposes, we also investigate the
performance of the solutions proposed in \cite{DetectionSymmetries}. Notice that these comparisons are performed in terms
of classification capabilities only, because the architectures proposed in \cite{DetectionSymmetries} are classifiers that
operate under $H_0$ of \eqref{eqn:MHP1}, namely, they select the most plausible PCM structure 
assuming that all the cells under considerations share the same covariance structure.

\begin{figure}[tbp] \centering
	\includegraphics[width=0.75\columnwidth]{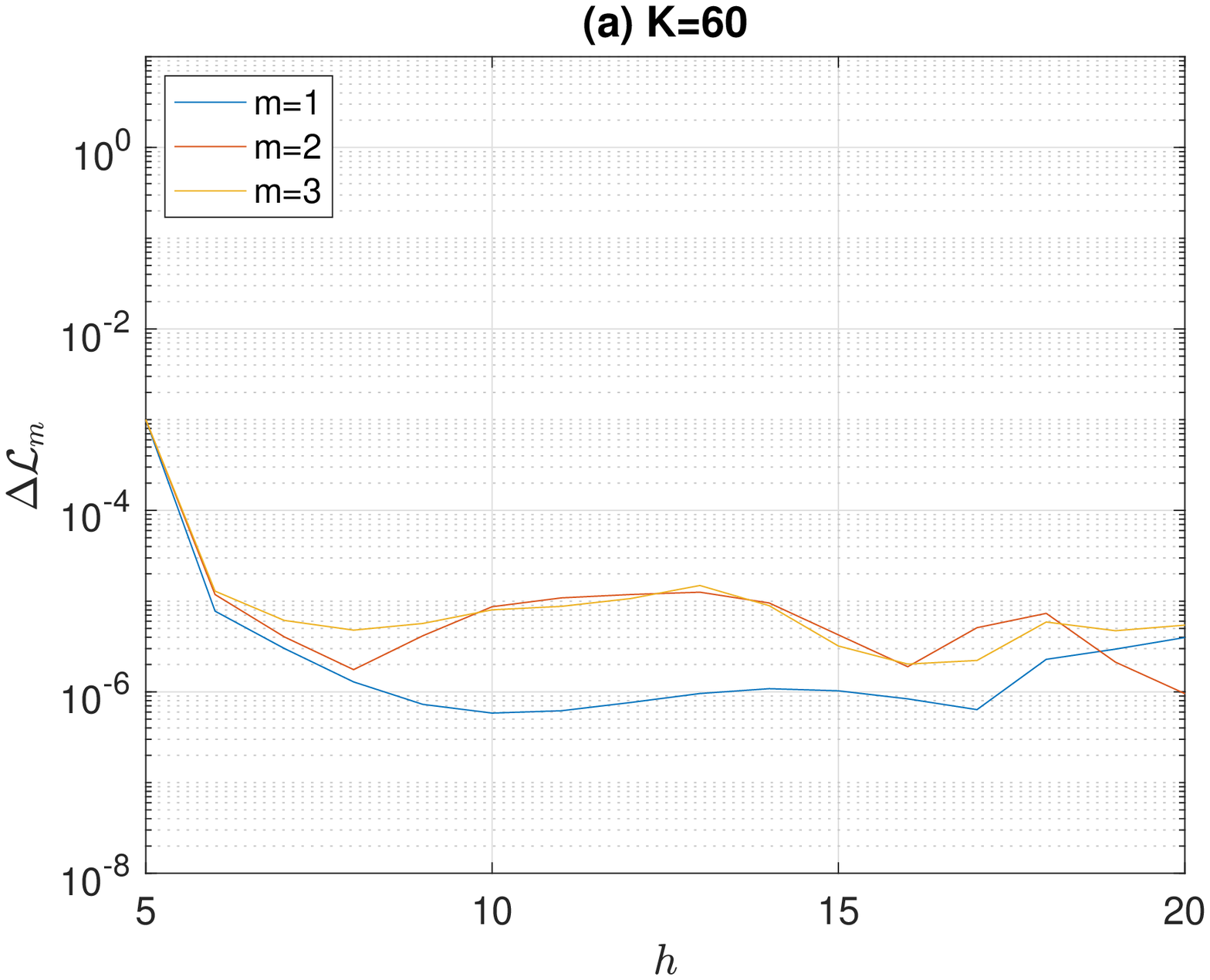}
\\
\vspace{3mm}
	\includegraphics[width=0.75\columnwidth]{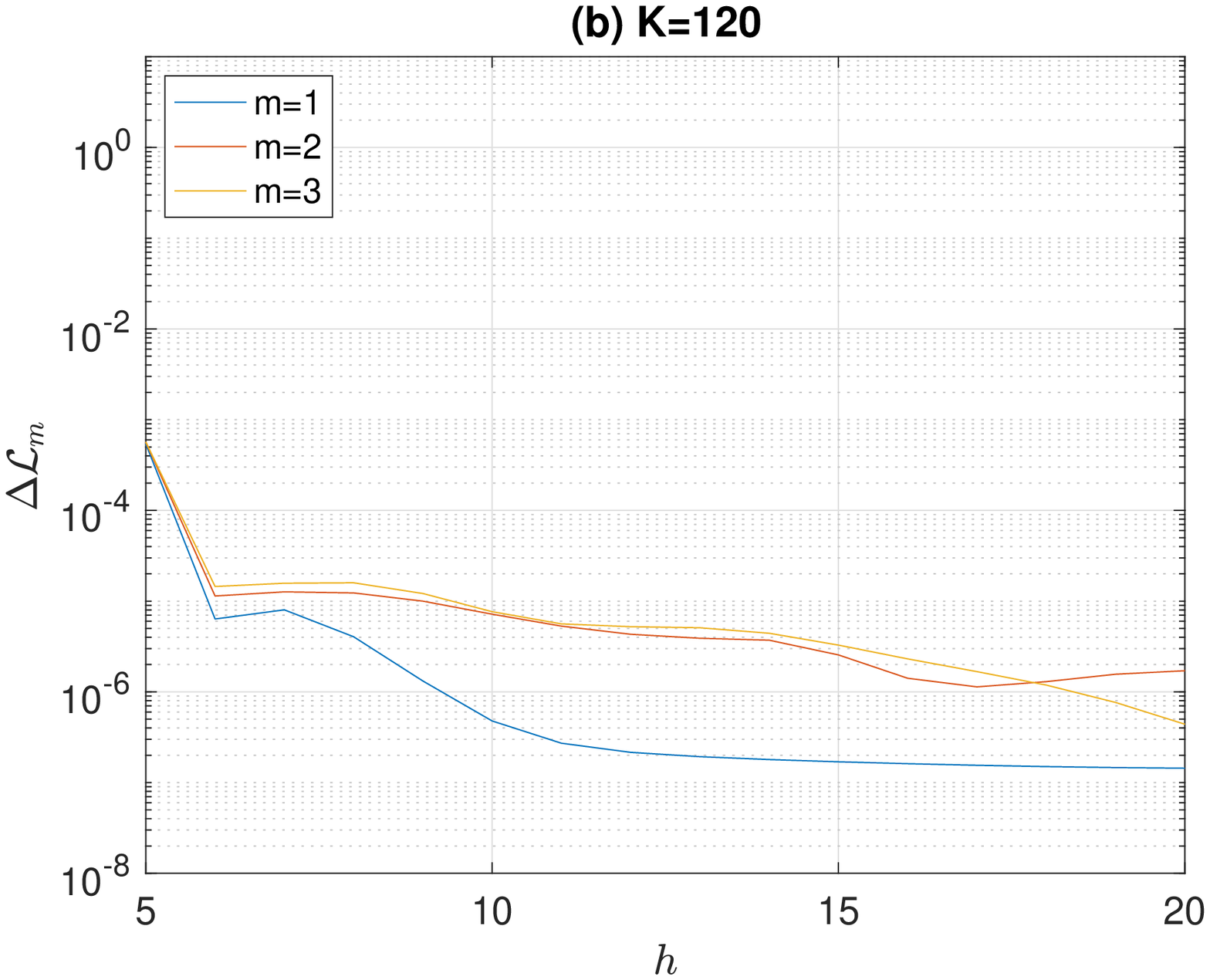}
\\
\vspace{3mm}
	\includegraphics[width=0.75\columnwidth]{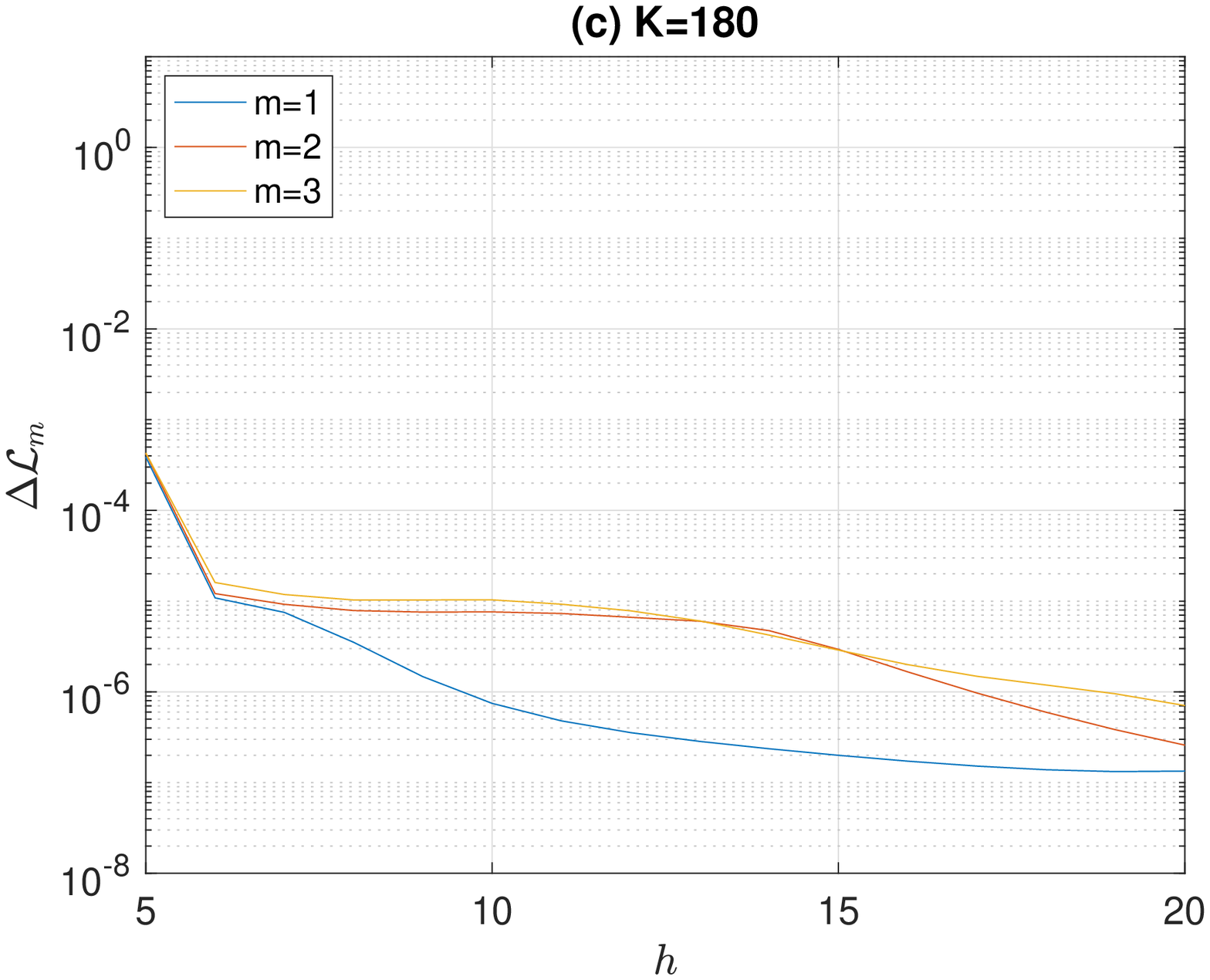}
\\
\vspace{3mm}
	\includegraphics[width=0.75\columnwidth]{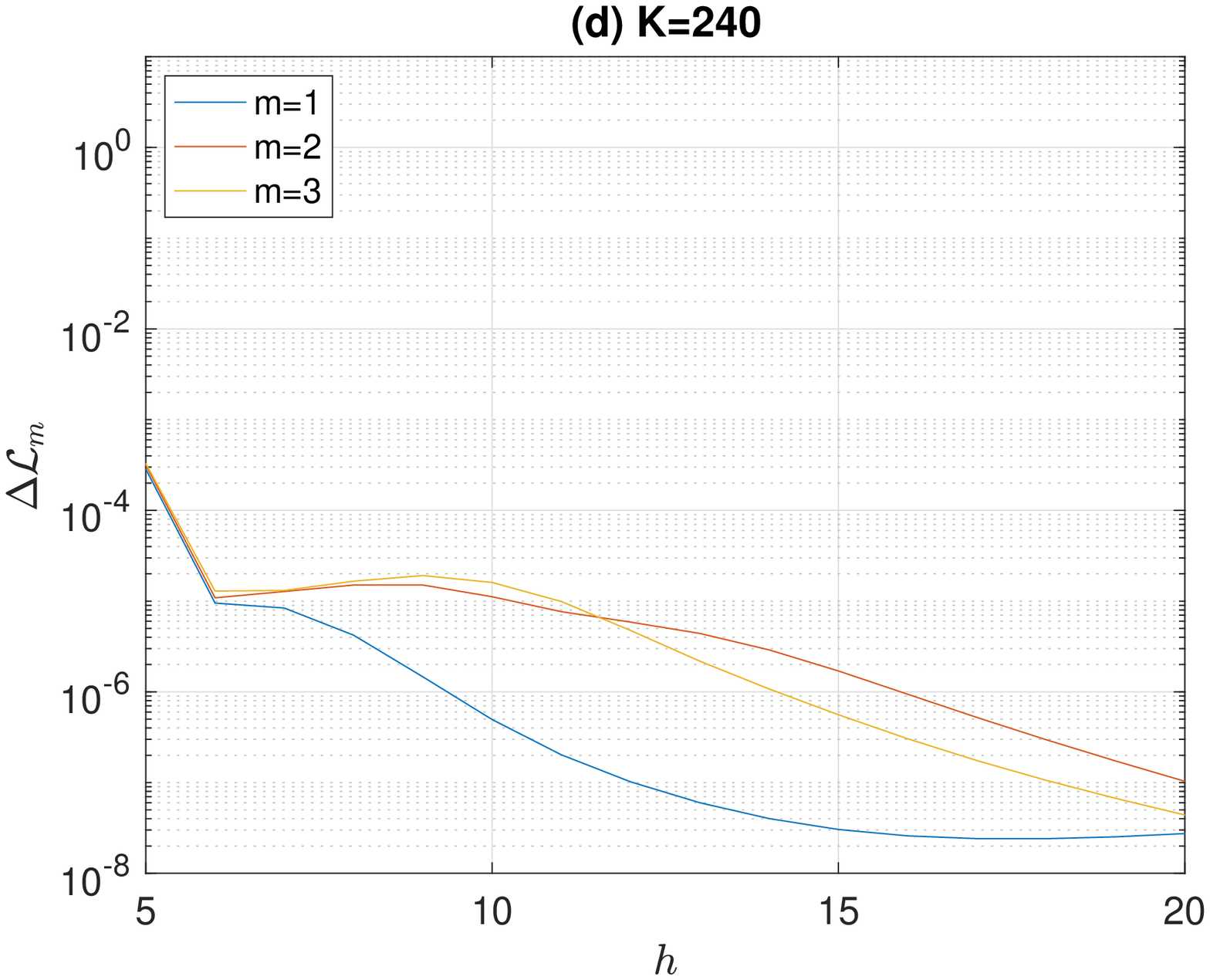}
\\
\vspace{3mm}
	\caption{Log-likelihood variations versus the iteration number $h$ of the EM algorithm for different values of $K$.}
	\label{figure_iterations}
\end{figure}

\begin{figure}[htb] \centering
	\subfigure[AIC-based]{\includegraphics[width=0.9\columnwidth]{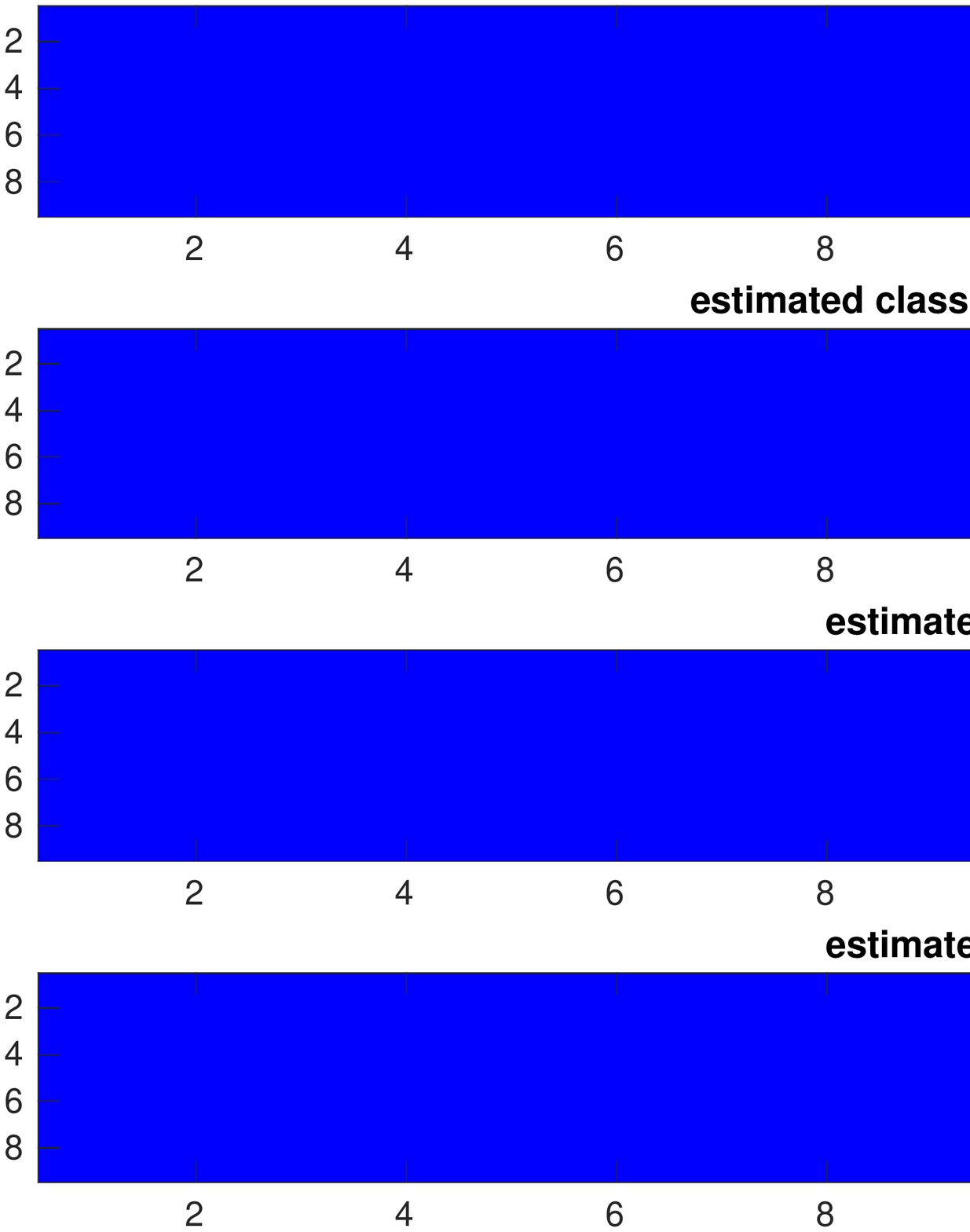}}
	\subfigure[BIC-based]{\includegraphics[width=0.9\columnwidth]{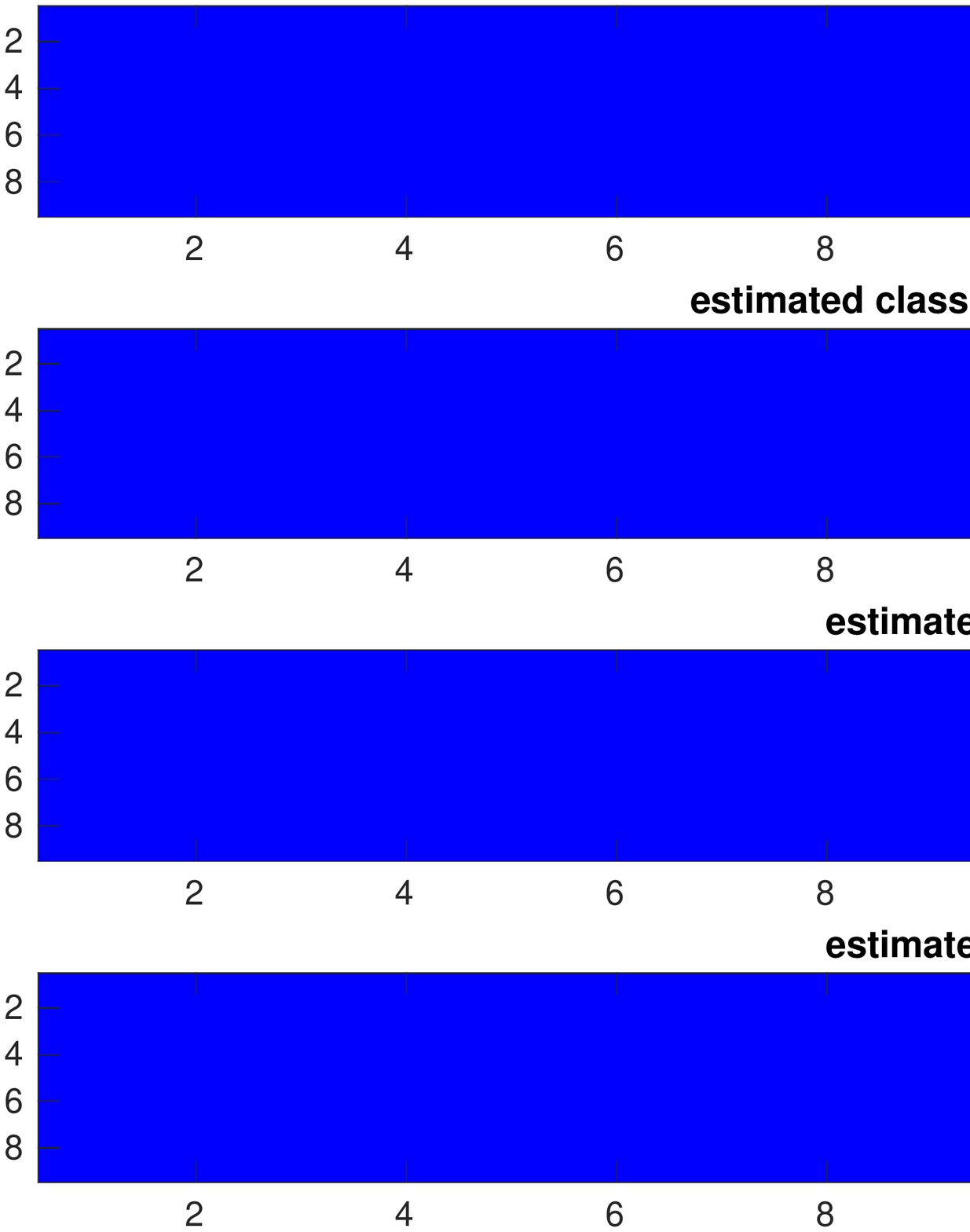}}
	\subfigure[GIC-based]{\includegraphics[width=0.9\columnwidth]{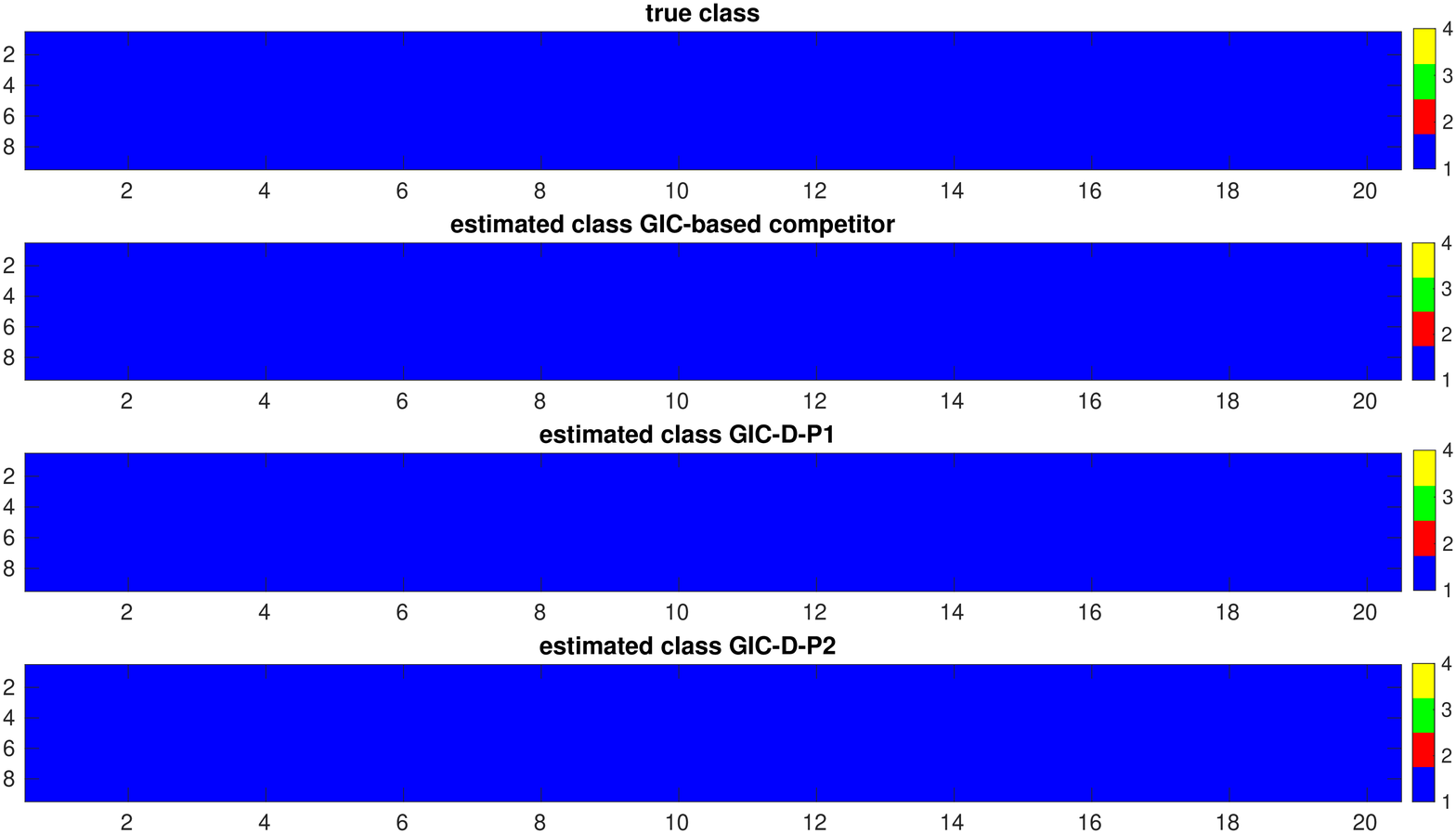}}
	\caption{Classification results for a single MC trial under $H_0$. }
	\label{figure_snapshots_H0}
\end{figure}

\begin{figure}[htb] \centering
	\subfigure[AIC-based]{\includegraphics[width=0.9\columnwidth]{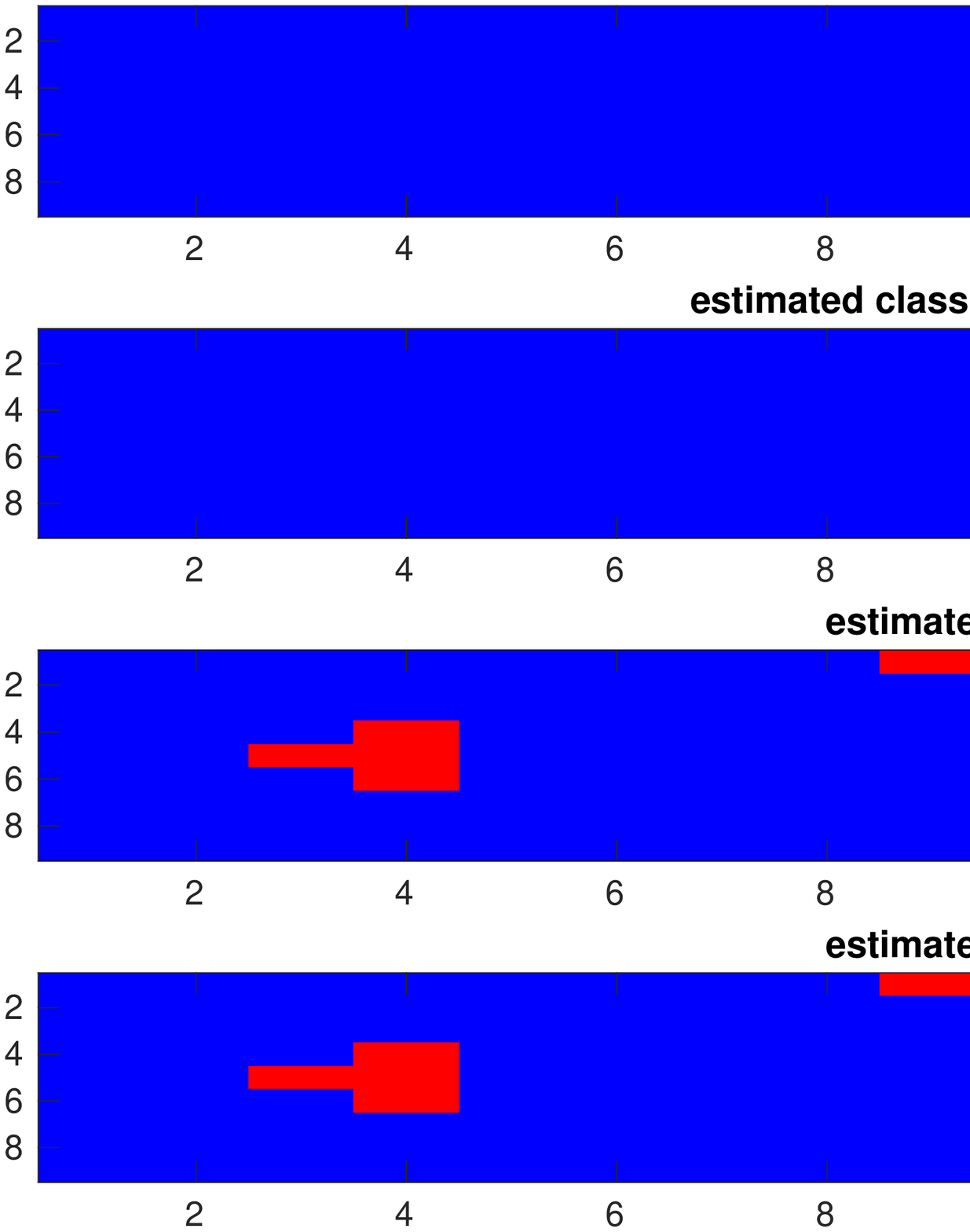}}
	\subfigure[BIC-based]{\includegraphics[width=0.9\columnwidth]{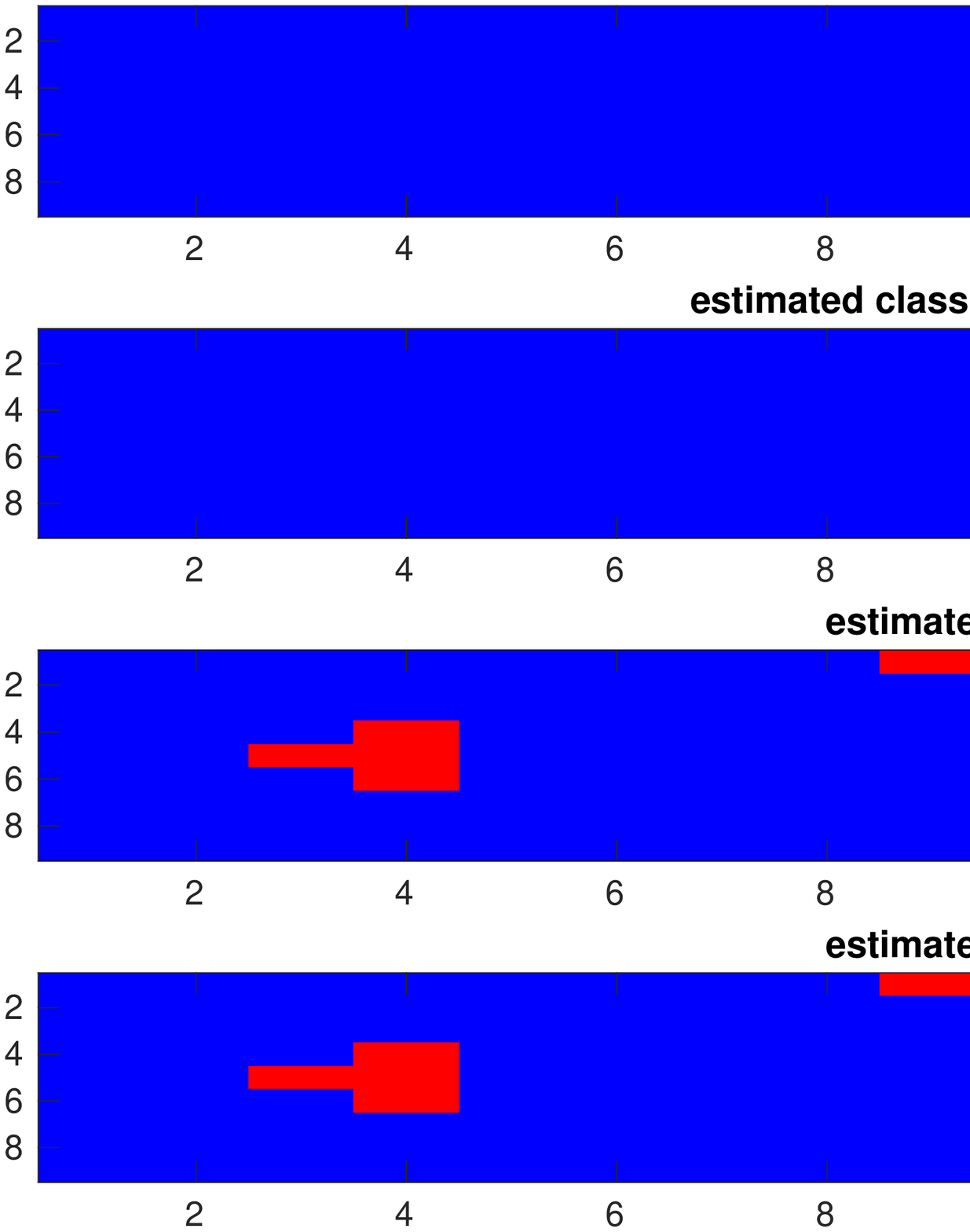}}
	\subfigure[GIC-based]{\includegraphics[width=0.9\columnwidth]{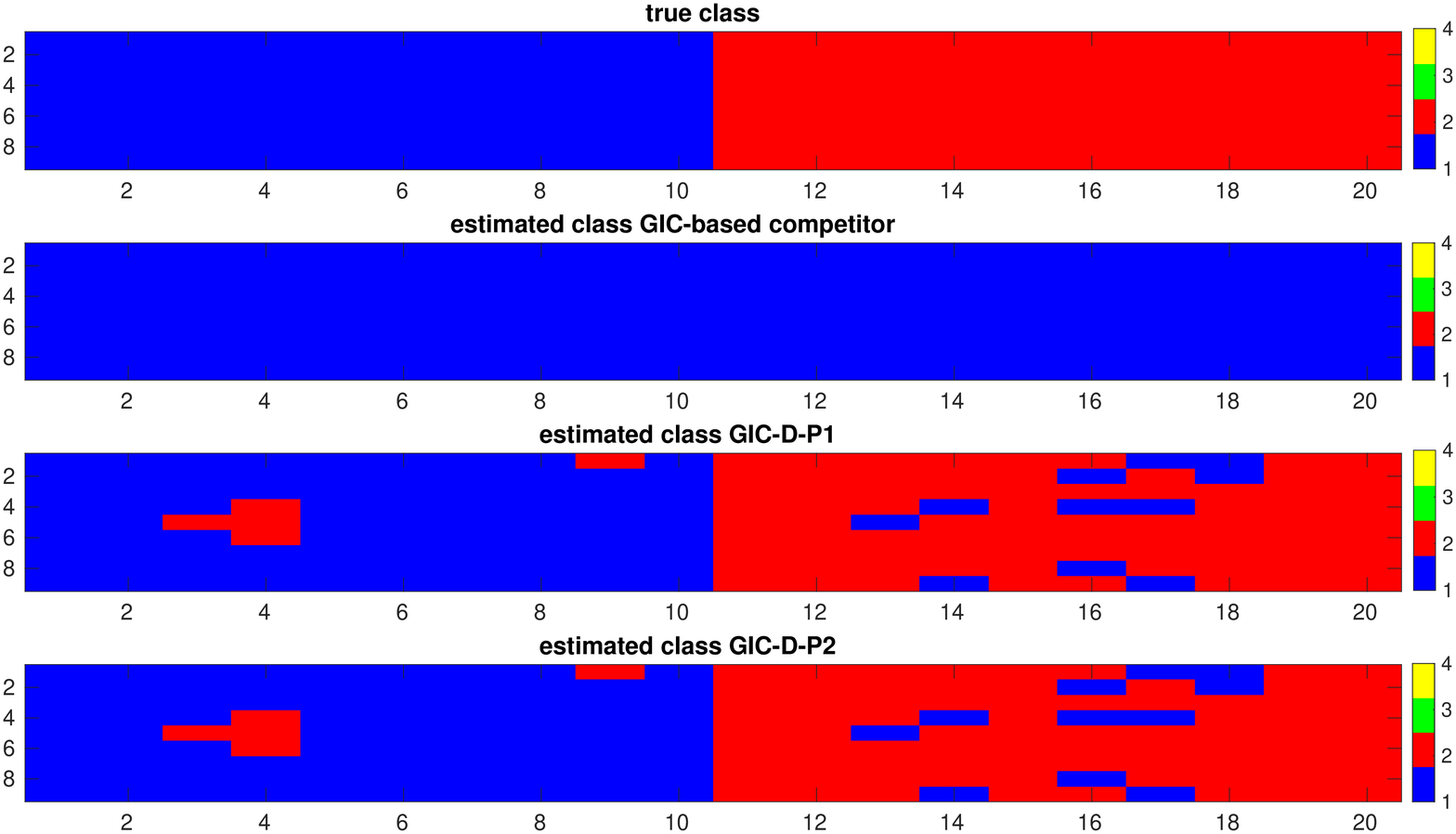}}
	\caption{Classification results for a single MC trial under $H_{1,1}$. }
	\label{figure_snapshots_H1}
\end{figure}

\begin{figure}[htb] \centering
	\subfigure[AIC-based]{\includegraphics[width=0.9\columnwidth]{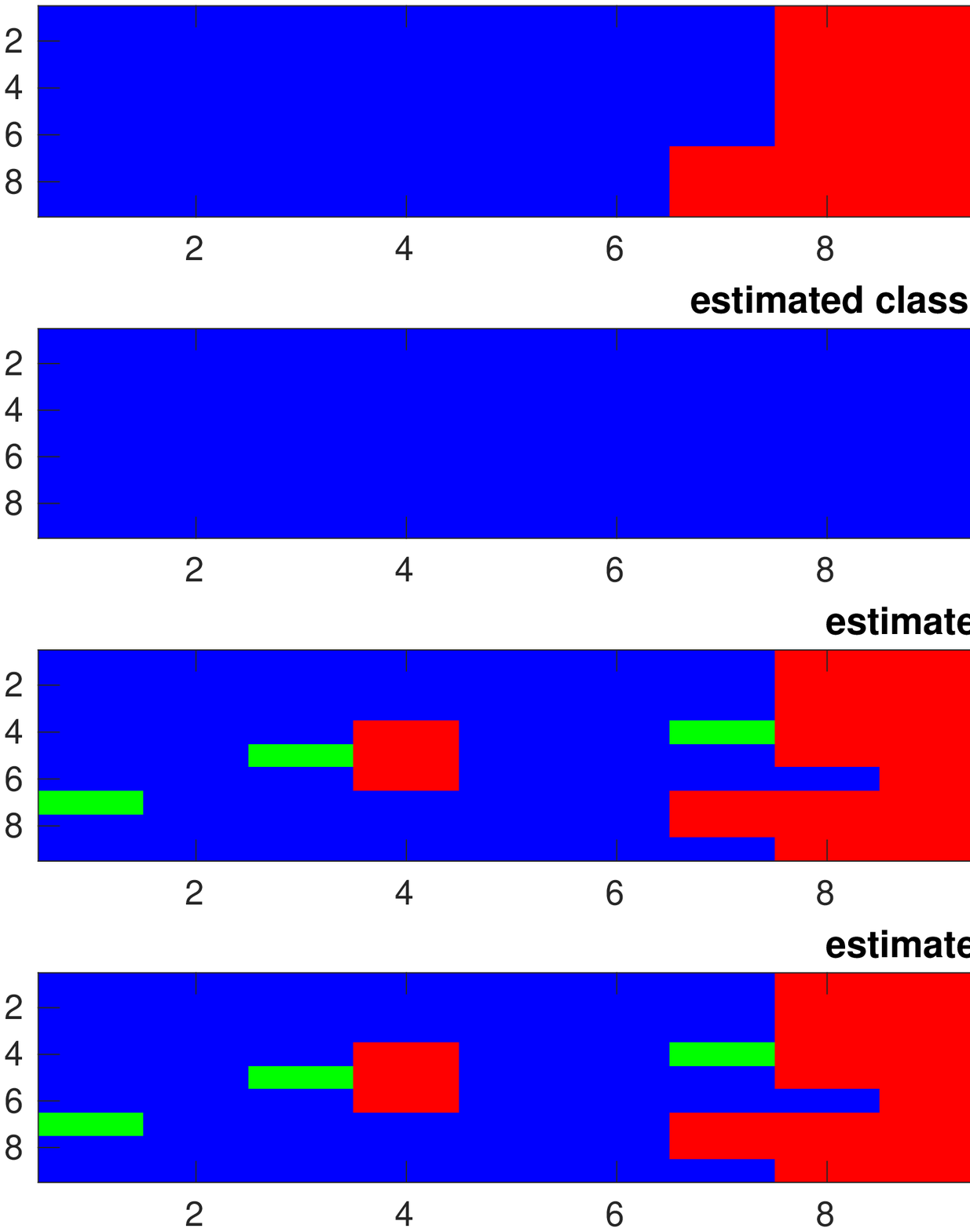}}
	\subfigure[BIC-based]{\includegraphics[width=0.9\columnwidth]{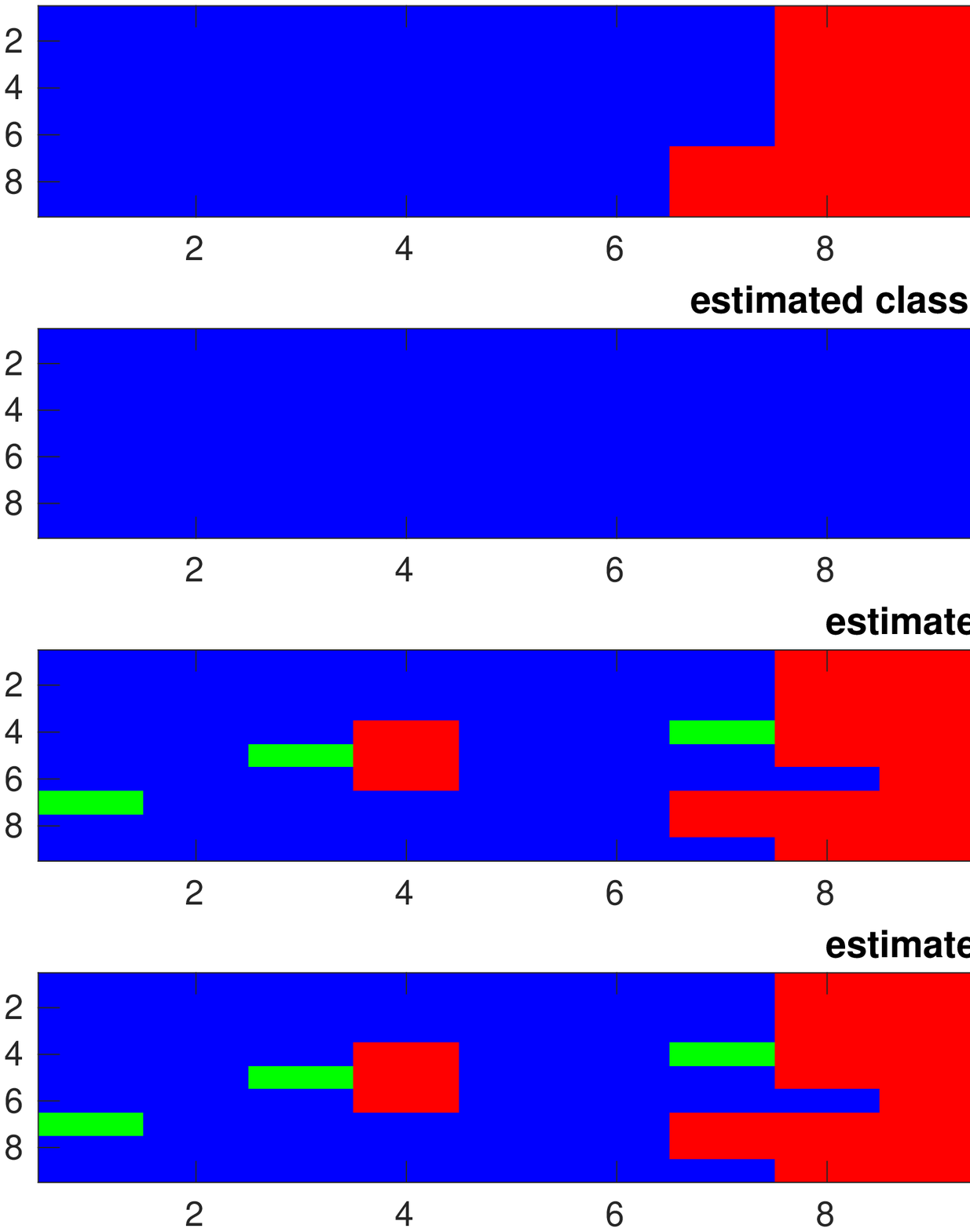}}
	\subfigure[GIC-based]{\includegraphics[width=0.9\columnwidth]{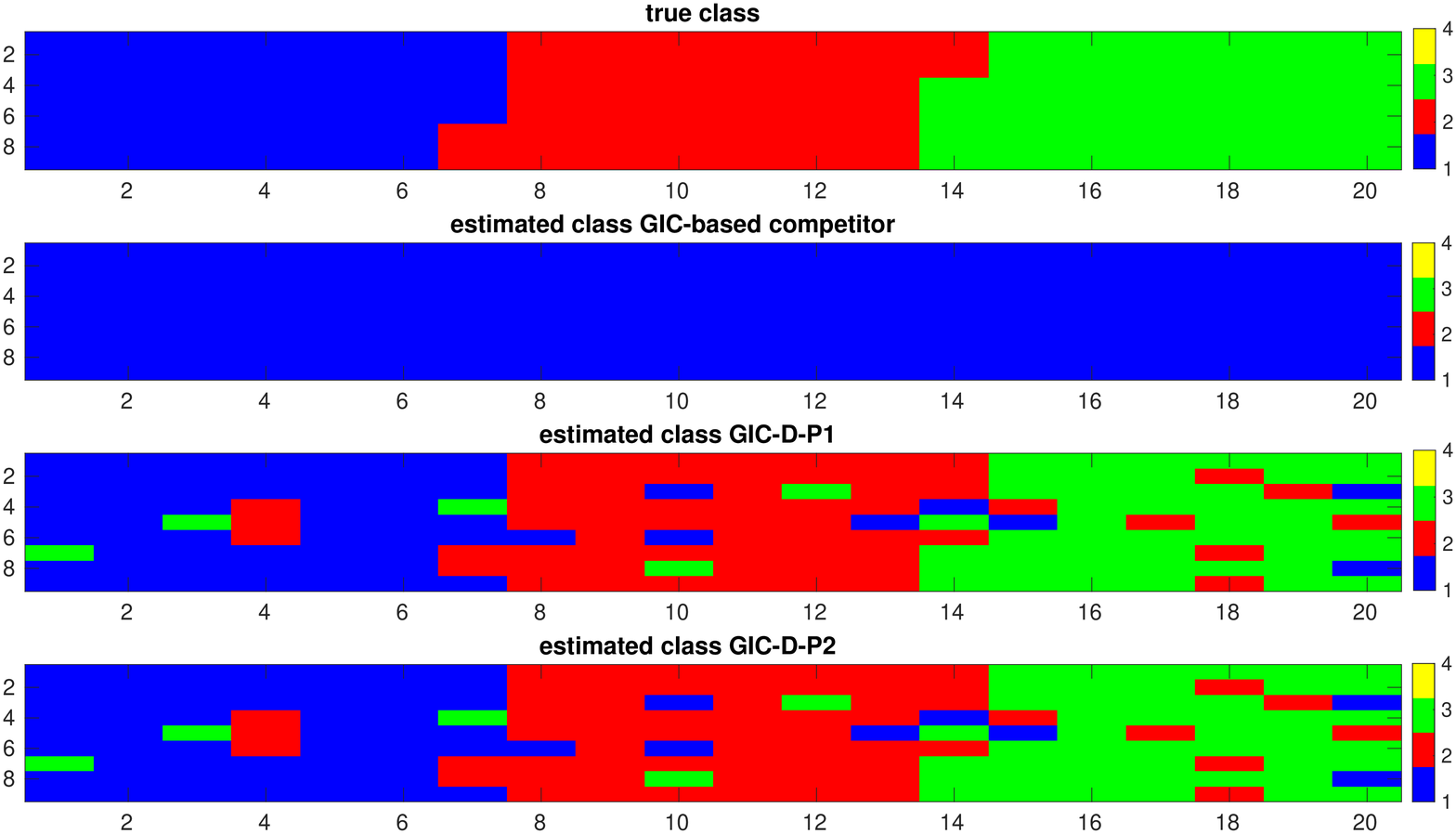}}
	\caption{Classification results for a single MC trial under $H_{1,2}$. }
	\label{figure_snapshots_H2}
\end{figure}

\begin{figure}[htb] \centering
	\subfigure[AIC-based]{\includegraphics[width=0.9\columnwidth]{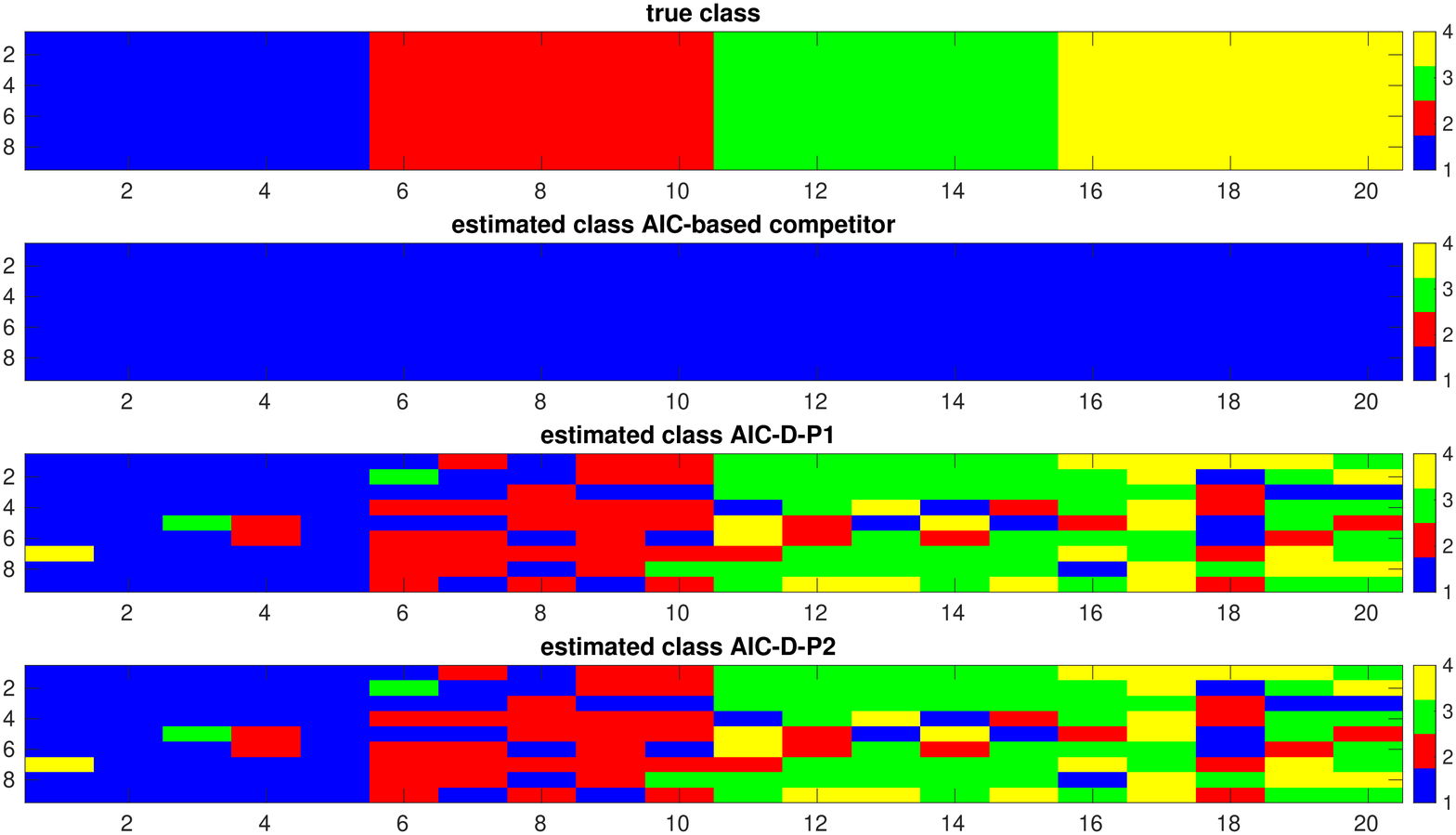}}
	\subfigure[BIC-based]{\includegraphics[width=0.9\columnwidth]{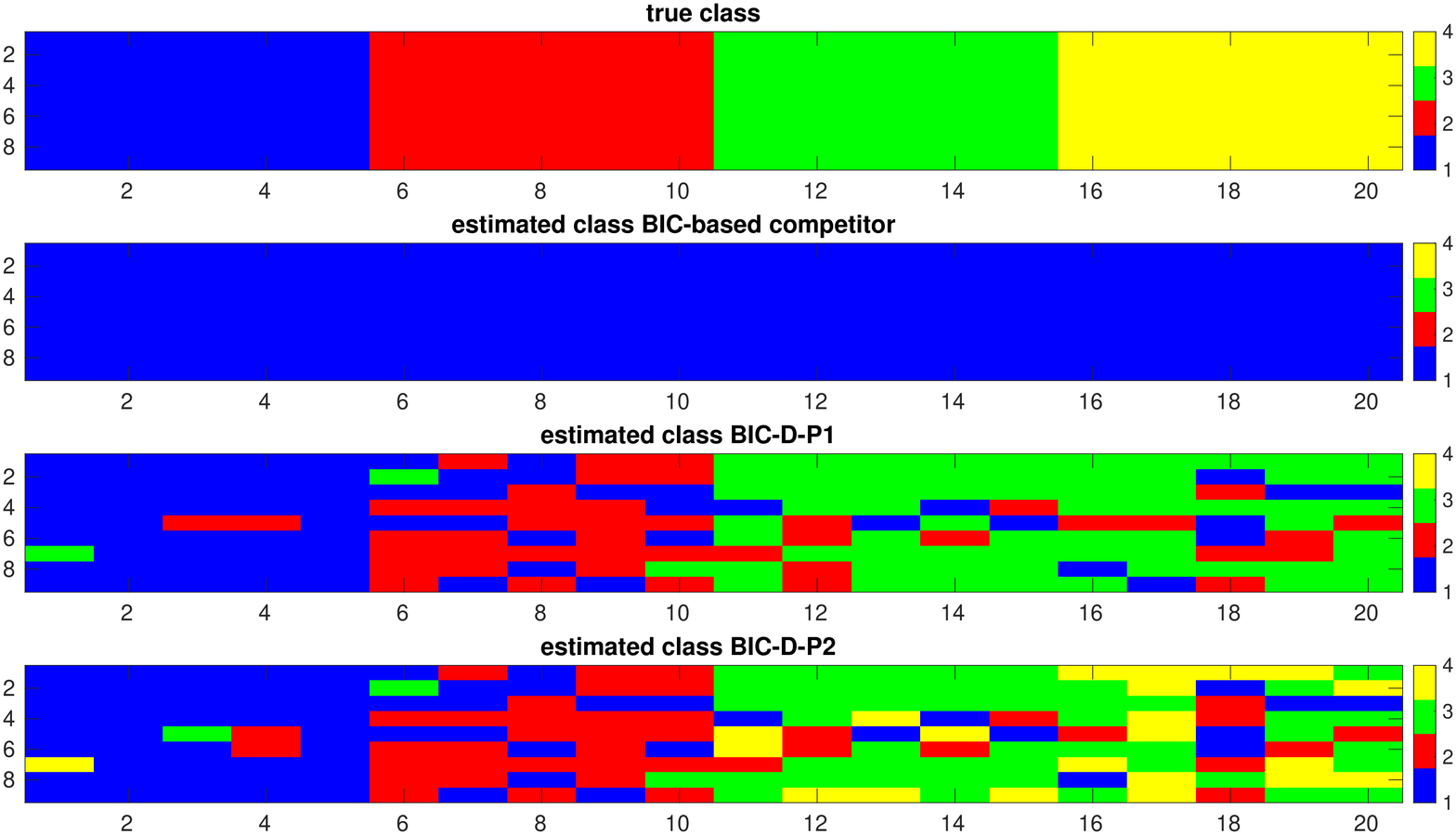}}
	\subfigure[GIC-based]{\includegraphics[width=0.9\columnwidth]{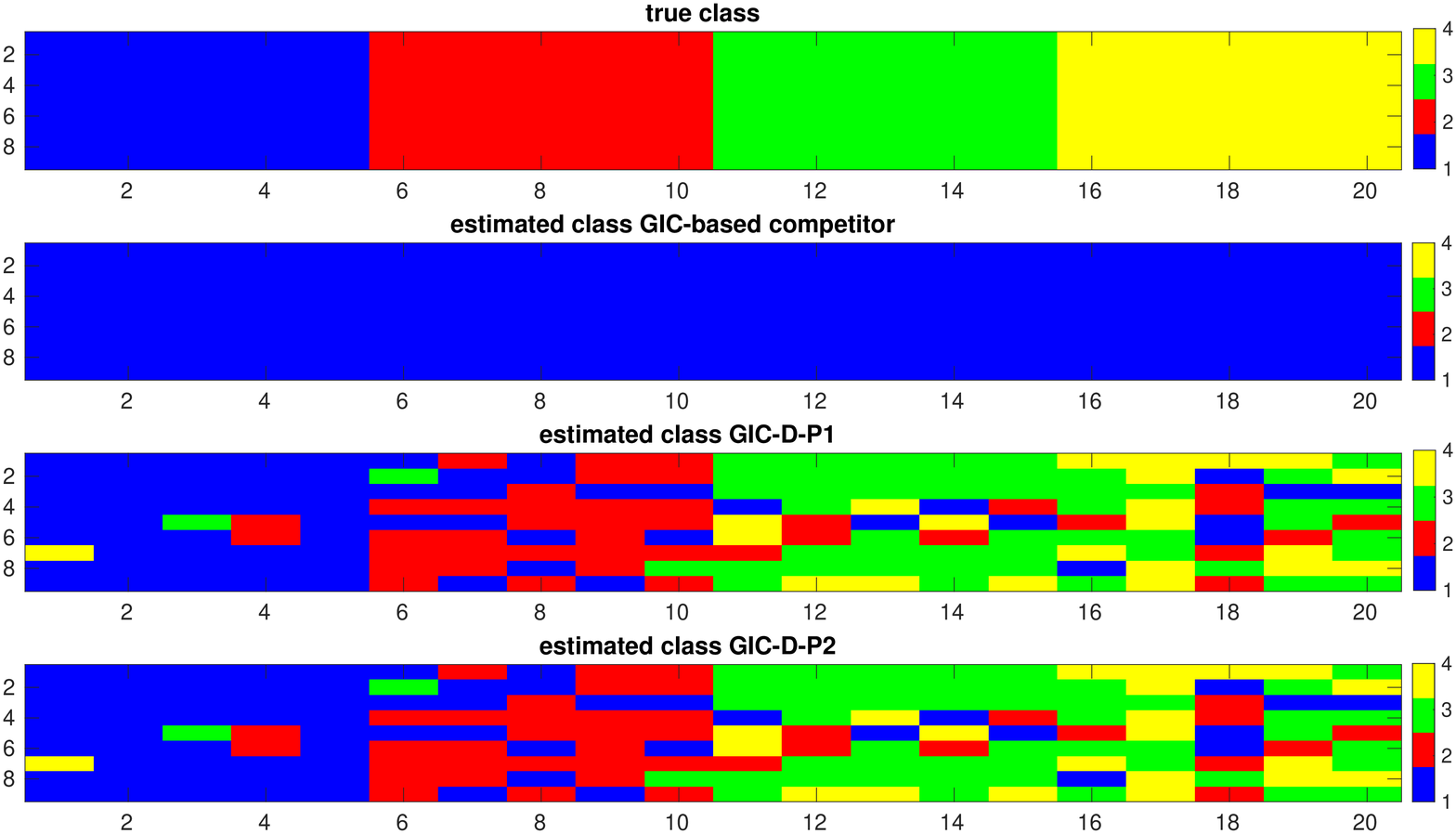}}
	\caption{Classification results for a single MC trial under $H_{1,3}$. }
	\label{figure_snapshots_H3}
\end{figure}

\begin{figure*}[htbp] \centering
{\includegraphics[width=2\columnwidth]{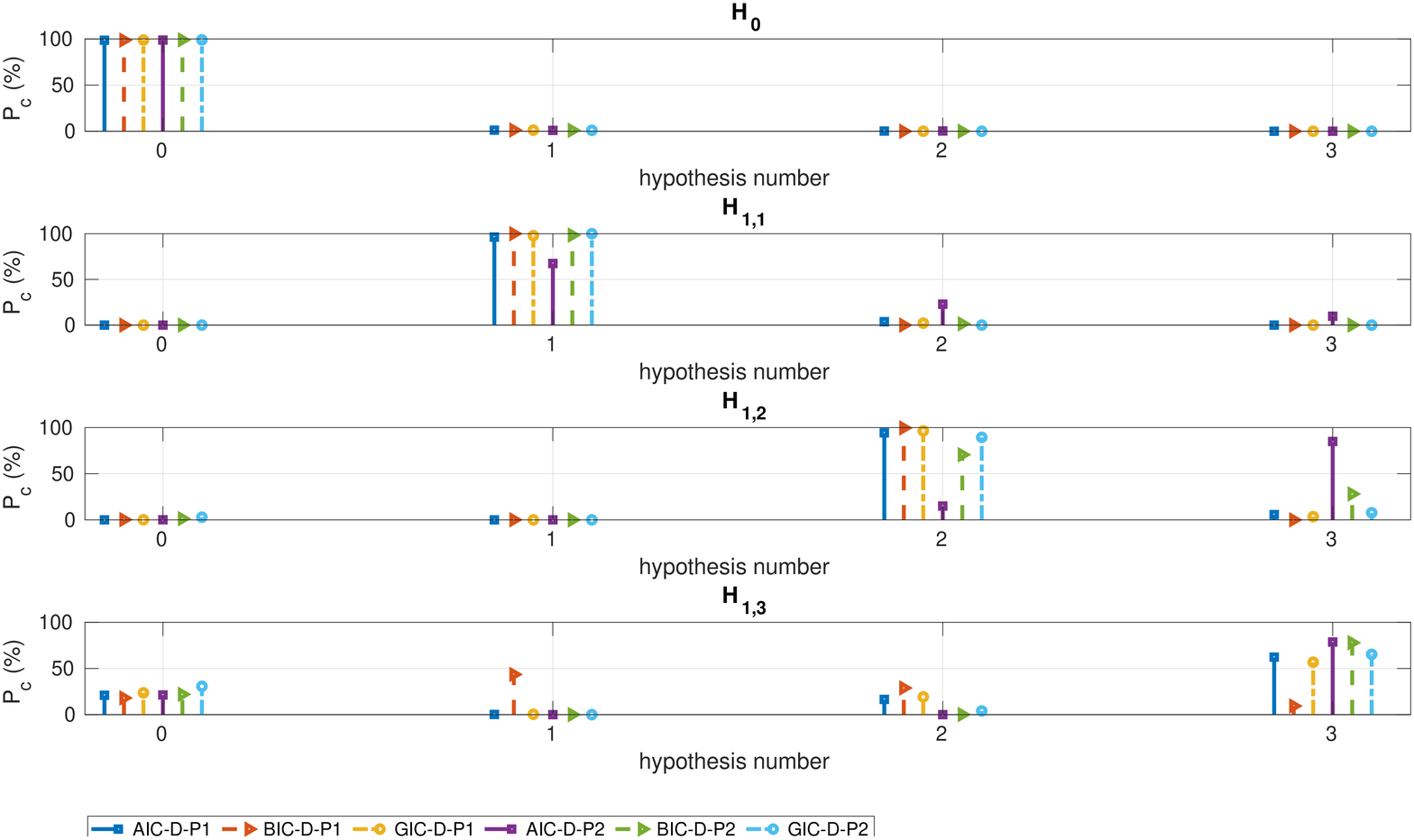}}
	\caption{$P_{c}$ ($\%$) versus $H_0$ and $H_{1,m}$, $m=1,2,3$, assuming that $K=120$.}
	\label{figurehisto120}
\end{figure*}

\begin{figure*}[htbp] \centering
{\includegraphics[width=2\columnwidth]{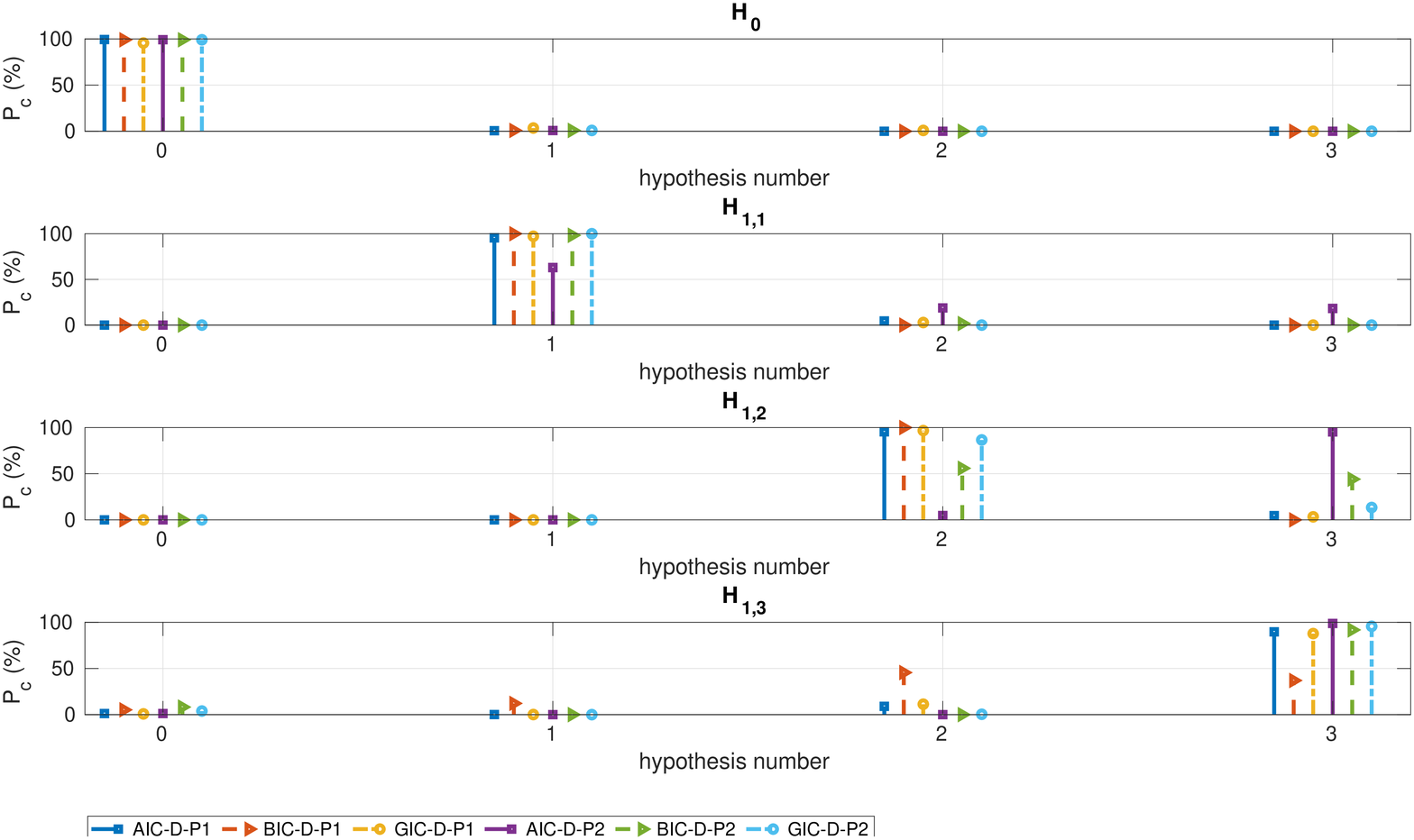}}
	\caption{$P_{c}$ ($\%$) versus $H_0$ and $H_{1,m}$, $m=1,2,3$, assuming that $K=180$.}
	\label{figurehisto180}
\end{figure*}

\begin{figure}[tbp] \centering
	\subfigure[$P_d$]{\includegraphics[width=0.75\columnwidth]{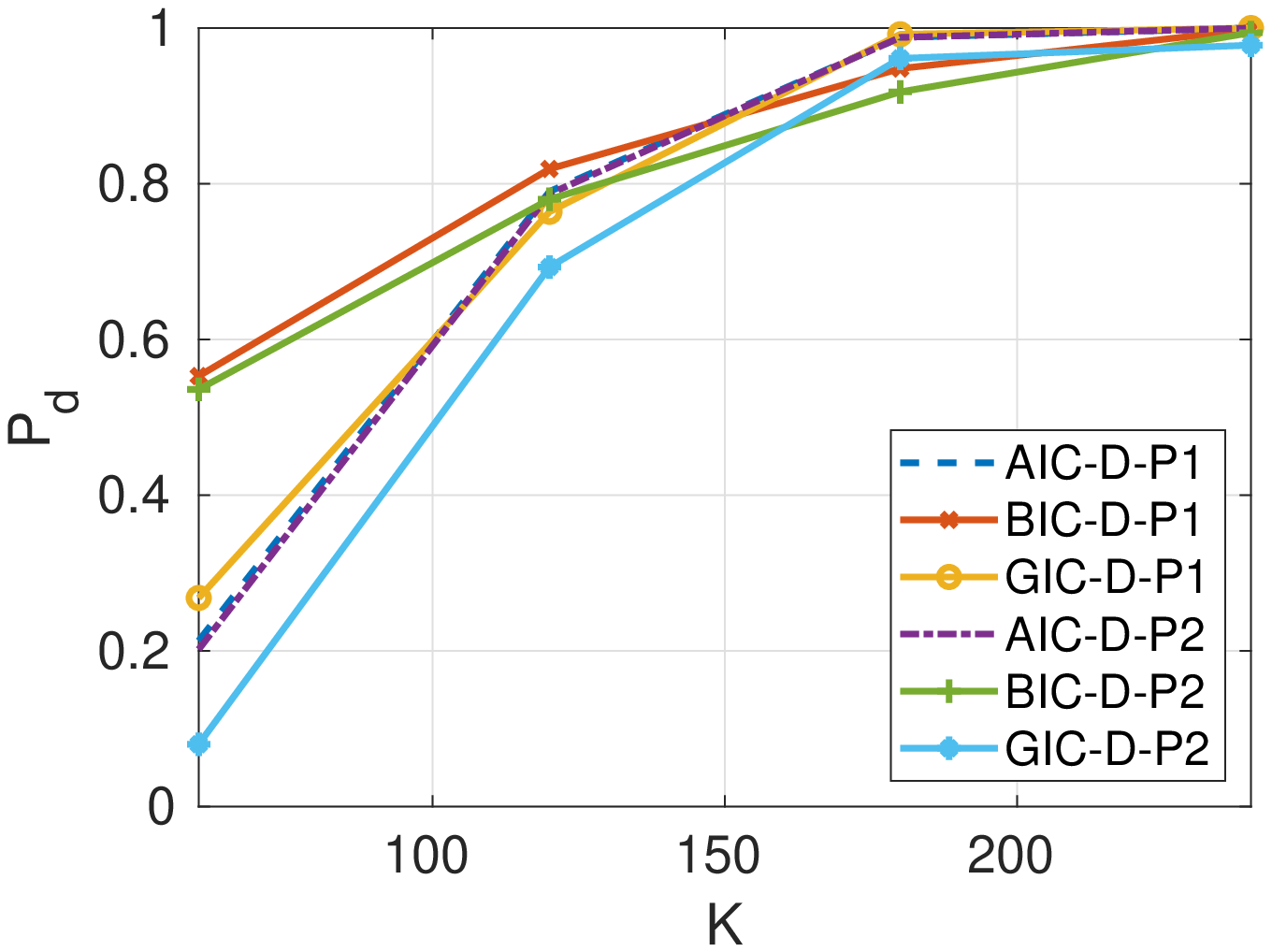}}
	\subfigure[RMSCE]{\includegraphics[width=0.75\columnwidth]{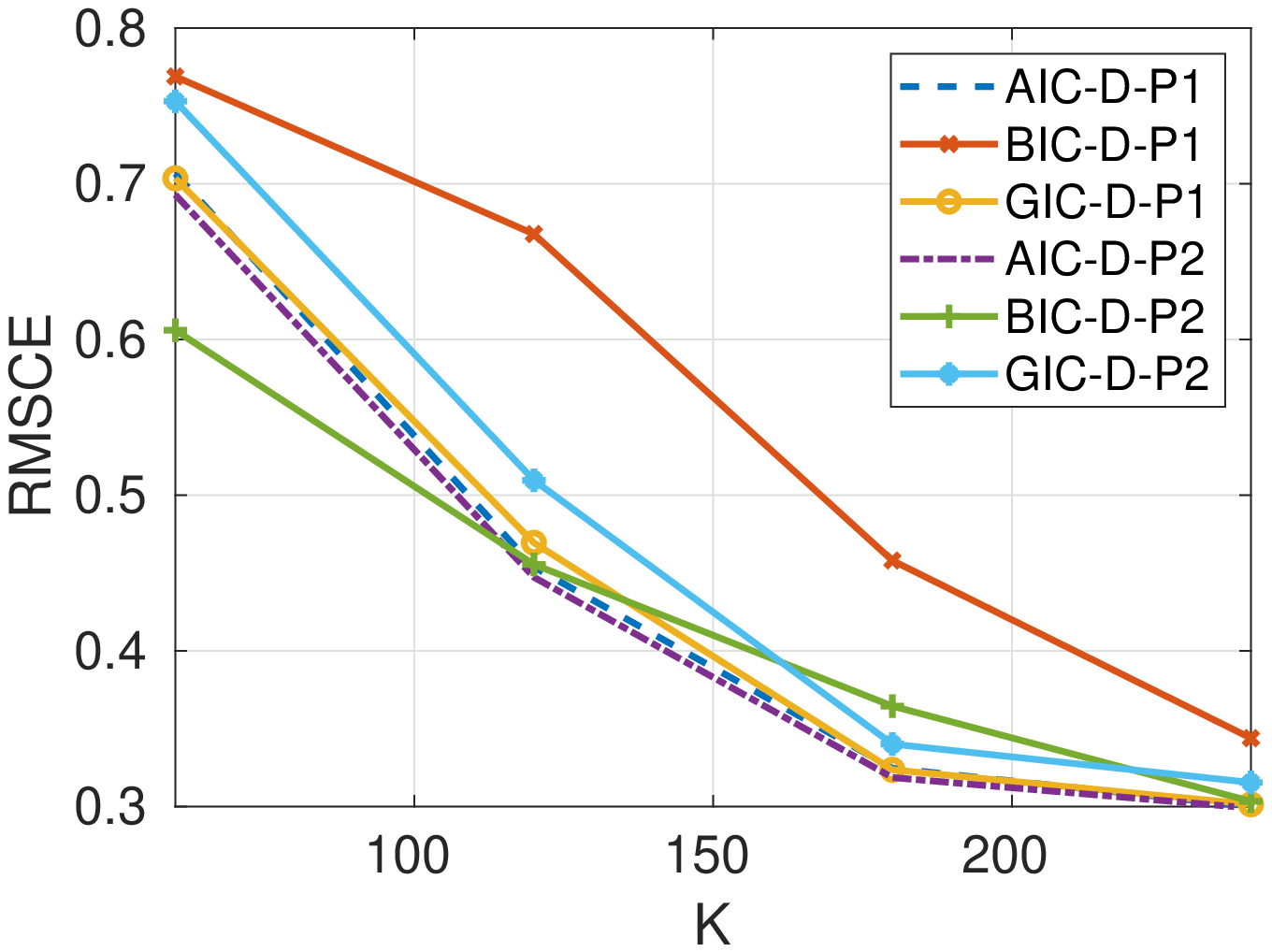}}
	\caption{$P_{d}$ and RMSCE versus $K$ assuming that $H_{1,3}$ is true.}
	\label{figure_Pd}
\end{figure}

\subsection{Simulated Data}

The simulated data obey the multivariate circular complex Gaussian distribution with zero mean 
and nominal covariance matrices related to four scenarios: no symmetry, reflection, rotation, and 
azimuth symmetries. Specifically, they are given by
\begin{align}
\bC_1 &=
\begin{bmatrix}
1 & 0.2+0.3j & 0.5-0.3j
\\
0.2-0.3j & 0.25 & -0.2-0.2j
\\
0.5+0.3j & -0.2+0.2j & 0.8
\end{bmatrix},
\\
\bC_2 &=
\begin{bmatrix}
1 & 0 & 0.5-0.3j
\\
0 & 0.25 & 0
\\
0.5+0.3j & 0 & 0.4
\end{bmatrix},
\\
\bC_3 &=
\begin{bmatrix}
1 & 0.3j & 0.2
\\
-0.3j & 0.4 & 0.3j
\\
0.2 & -0.3j & 1
\end{bmatrix},
\\
\bC_4 &=
\begin{bmatrix}
1 & 0 & 0.5
\\
0 & 0.25 & 0
\\
0.5 & 0 & 1
\end{bmatrix},
\end{align}
respectively. In the numerical examples below, the number of data ($K$) ranges from $60$ to $240$, and data are partitioned 
into adjacent subsets characterized by different PCM structures. 
The parameter $\rho$ (of GIC-based architectures) is set to $3$ for the competitor \cite{DetectionSymmetries}, 
$1.3$ for GIC-D-P1, and $11$ for GIC-D-P2 (these values are selected in order to guarantee
a good compromise between underestimation and overestimation of the model order).
Finally, we consider $P_{fa} =10^{-2}$ and the related detection thresholds are estimated as follows
\begin{enumerate}
\item compute the detection threshold under $H_0$ and for each PCM structure;
\item the final threshold (namely, $\eta$ in \eqref{eqn:penalizedLLRT}) is set by selecting the maximum of the 
thresholds obtained at the previous step.
\end{enumerate}
The above procedure guarantees that the actual $P_{fa}$ is less than or equal to the nominal $P_{fa}$.

As a preliminary analysis, we focus on the requirements of the proposed procedures in terms of the EM iterations. 
To this end, in Figure \ref{figure_iterations}, we plot the log-likelihood variations, i.e., $\Delta \cL_m(h)$, $m=1,2,3$, 
as a function of $h$, averaged over 1000 MC trials. It turns out that, for all the analyzed cases, a number of 10 iterations 
(this value will be used in the subsequent analysis) is sufficient to ensure log-likelihood variations less 
than $10^{-4}$, namely, $\epsilon_m<10^{-4}$. 

In Figures \ref{figure_snapshots_H0}-\ref{figure_snapshots_H3}, we investigate the instantaneous behavior of the proposed
architectures by showing the classification outcomes of a single 
Monte Carlo trial using a window of size $9\times20$. These figures are obtained by generating data as follows
\begin{itemize}
\item under $H_0$, all data share $\bC_1$;
\item under $H_{1,1}$, data are split in two equal parts, where the PCM of the first and second halves are $\bC_1$ and $\bC_2$, respectively; 
\item under $H_{1,2}$, data are partitioned into three subsets with the same cardinality and 
characterized by $\bC_1$, $\bC_2$, and $\bC_3$;
\item under $H_{1,3}$, four equal subsets are generated and, clearly, all the PCMs are used.
\end{itemize}
In these figures, the estimated structure is mapped to its structure index, namely, $i$ ($\in\{1,\ldots,4\}$) means that $\bC_i$ has
been selected. The ground truth is reported at the beginning of each subfigure.
From the figures' inspection, it is evident the advantage (at least from a qualitative point of view) of 
the proposed architectures over the considered competitor \cite{DetectionSymmetries}, 
when $H_{1,m}, m=1,\dots,3$, is in force. 
Moreover, as expected, by comparing the four figures, it is possible to observe that $H_{1,3}$ represents 
the most challenging scenario with the major difficulties in correctly classifying the azimuth symmetry 
(yellow pixels present in the last partition of data set). As a matter of fact,
AIC-D-P1, AIC-D-P2, BIC-D-P2, GIC-D-P1, and GIC-D-P2 are capable of only partially classifying such pixels as characterized
by azimuth symmetry.

A more quantitative analysis is performed in Figures \ref{figurehisto120} and \ref{figurehisto180} 
that show the histograms of correct classification over $1000$ independent MC trials assuming
$K=120$ and $K=180$, respectively. 
Such histograms are representative of the probability
of correct classification ($P_c$) defined as the probability of declaring $H_0$ or $H_{1,m}$, $m=1,\dots,3$, under $H_0$ 
or $H_{1,m}$, respectively.
As expected, under $H_0$, all the proposed architectures return $P_{c}$ values very close to 100$\%$. 
Under $H_{1,1}$, all the considered architectures can provide percentages of correct classification close to 100$\%$
except for AIC-D-P2 whose $P_c$ values are around $0.70$.
Almost similar behaviors can be observed under $H_{1,2}$ with the difference that architectures based upon the second 
EM-based procedure have lower classification capabilities with respect to the results under $H_{1,1}$.
Under this hypothesis, the performance of AIC-D-P2 is very poor due to a strong overestimation inclination.
Such inclination is also experienced by BIC-D-P2 for $K=180$ since the resulting $P_c$ is about $0.56$.
Under $H_{1,3}$, which represents the most challenging case, we notice that for $K=120$ the classification values
are below $0.75$ for all the considered architectures with BIC-D-P1 returning the worst performance.
When $K$ increases to $180$, the situation is clearly better than for $K=120$ even though the classification performance
of BIC-D-P1 is less than $0.5$. The other architectures ensure $P_{c}$ values greater than 92$\%$. 

The curves reported in Figure \ref{figure_Pd} pertain to the probability of PCM variation detection ($P_{d}$) 
and the normalized root mean square classification error (RMSCE) values both as functions of $K$; 
notice that the $P_{d}$ is defined as the probability of rejecting $H_0$ under $H_{1,m}$, whereas
the RMSCE is the root mean square number of misclassified vectors divided\footnote{This normalization
is necessary for comparison purposes.} by $K$. 
Data are generated under the most challenging hypothesis $H_{1,3}$ and, again, the performance parameters 
are estimated over $1000$ MC independent trials.
From Subfigure \ref{figure_Pd}(a), it turns out that the curves associated with the considered architectures are close 
to each other when $K>120$ with a maximum difference of about $0.1$. This difference becomes negligible
as $K$ increases. As a matter of fact, AIC-D-P1, AIC-D-P2, and GIC-D-P1 are capable of achieving 
$P_d=1$ at $K=240$, whereas BIC-D-P1, BIC-D-P2, and GIC-D-P2 return $P_d=0.998$, $P_d=0.994$, and $P_d=0.978$, 
respectively, at $K=240$. In Subfigure \ref{figure_Pd}(b), we plot the normalized RMSCE versus $K$. 
The figure points out that the error curves for AIC-D-P1, AIC-D-P2, and GIC-D-P1 are almost overlapped
outperforming the other classifiers at least for $K<240$. The worst performance is returned by
BIC-D-P1 as expected from the analysis of the classification histograms.

Summarizing, the above analysis indicates that AIC-D-P1, GIC-D-P1, and AIC-D-P2
can guarantee an excellent compromise between detection performance and classification results
under each hypothesis for $K>120$. 
In addition, notice that if we consider subsets of hypotheses, other architectures can provide
reliable classification and detection performance starting from $K=120$.

\subsection{Real Recorded Data}
\label{subsec:realData}
\begin{figure}[tbp]
    \centering
    \includegraphics[width=.45\textwidth]{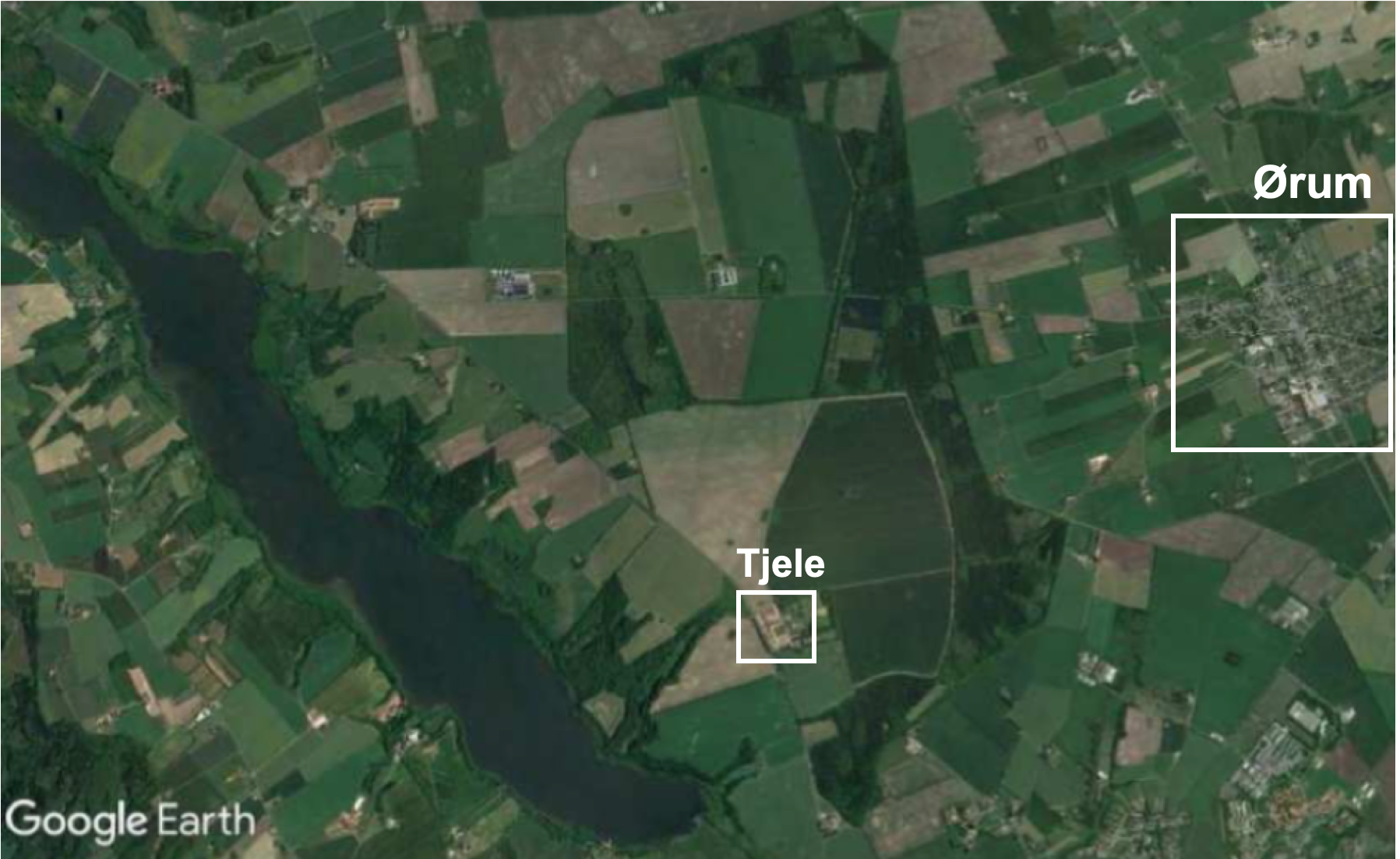}
    \caption{Optical image of the observed scene (drawn from Google Earth \copyright).}
    \label{fig:optical_real}
\end{figure}

In this last subsection, we consider the fully polarimetric SAR data acquired by the EMISAR airborne 
sensor\footnote{Data can be downloaded at: https://earth.esa.int/web/polsarpro/data-sources/sampledatasets.} 
in the L-band (1.25 GHz). The set is formed by 1750 rows and 1000 columns. 
The scene under investigation is over the Foulum Area, Denmark, and contains a mixed urban, vegetation, 
as well as water scene (as shown in Figure \ref{fig:optical_real}).
Therefore, it is representative of different scattering mechanisms that allow us to suitably verify the classification capabilities
of the proposed algorithms in a real-world manifold scenario.
The rectangular boxes in the figure highlight the two urban areas of Tjele and Orum.

\begin{figure*}[tbp] \centering
\subfigure[AIC-based competitor]{\includegraphics[width=0.6\columnwidth]{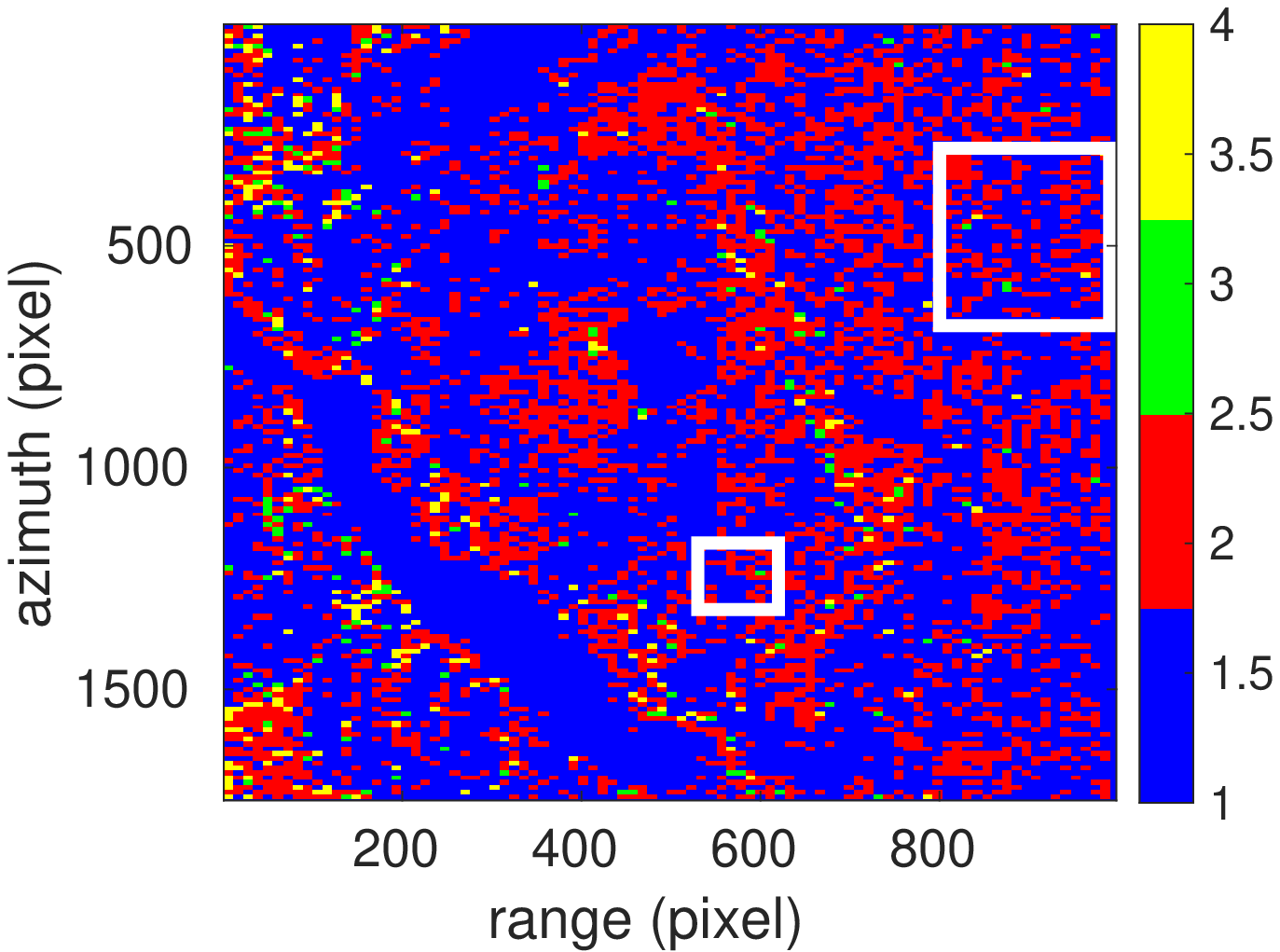}}
	\subfigure[AIC-D-P1]{\includegraphics[width=0.6\columnwidth]{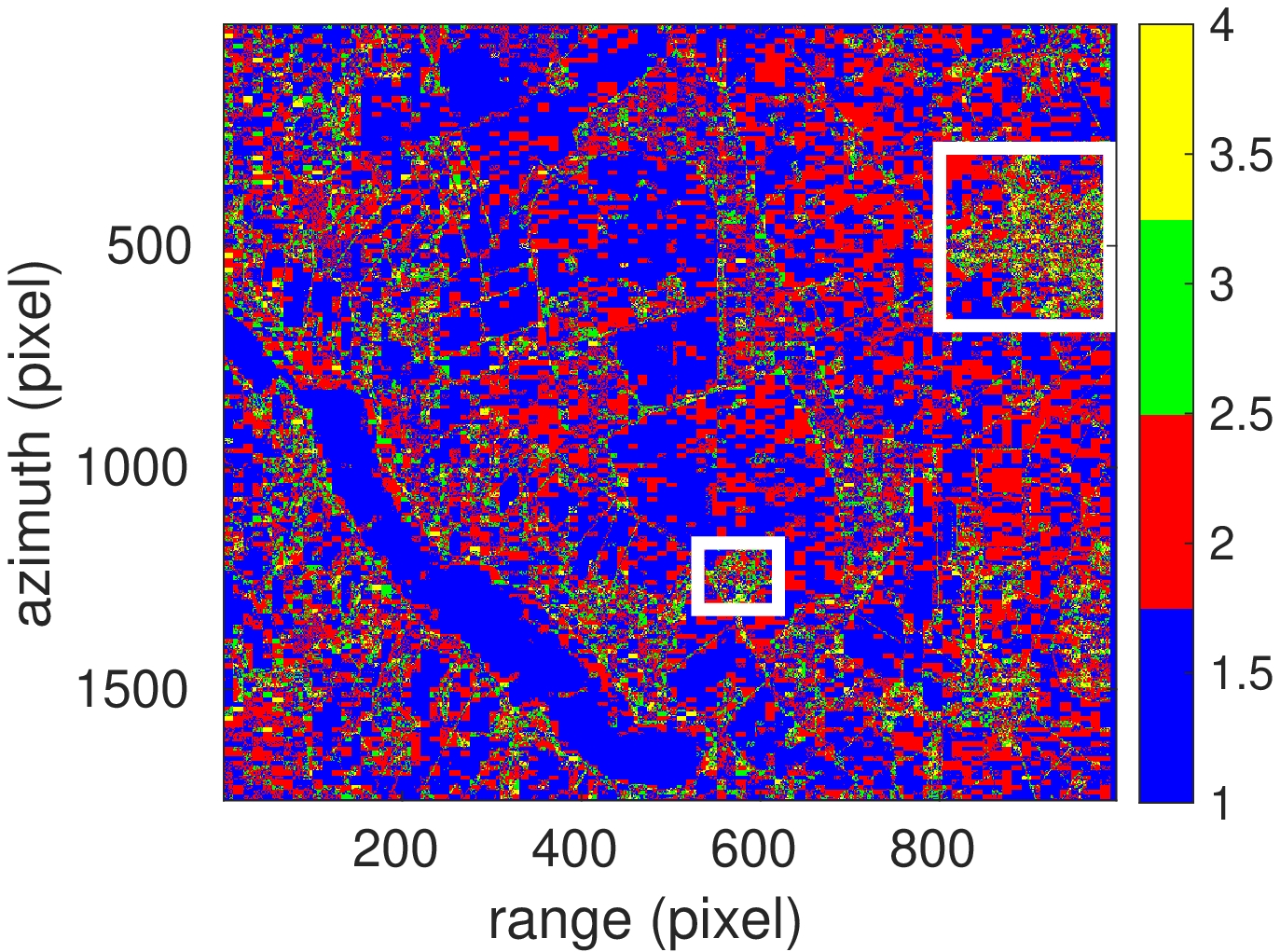}}
		\subfigure[AIC-D-P2]{\includegraphics[width=0.6\columnwidth]{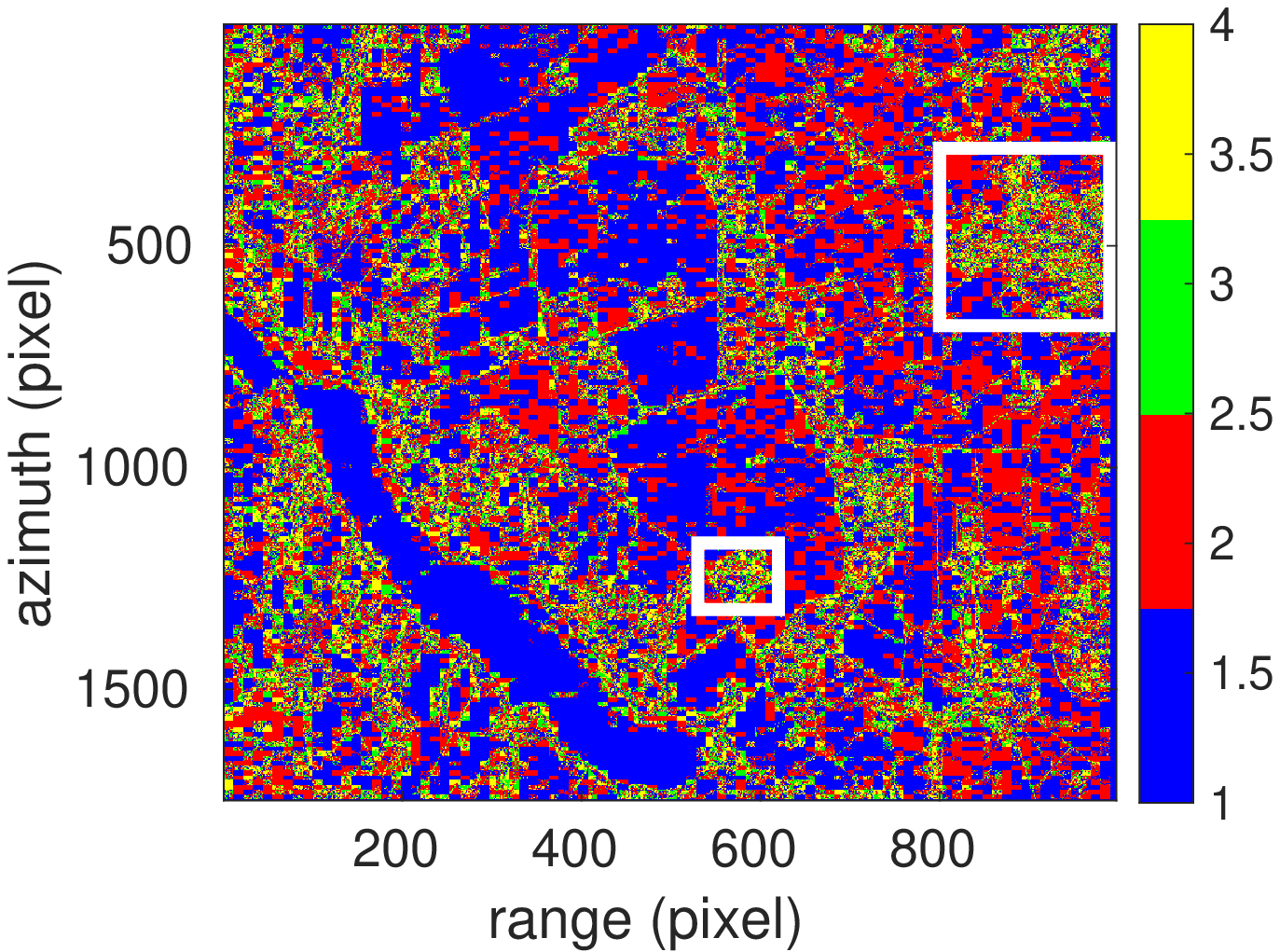}}
		\subfigure[BIC-based competitor]{\includegraphics[width=0.6\columnwidth]{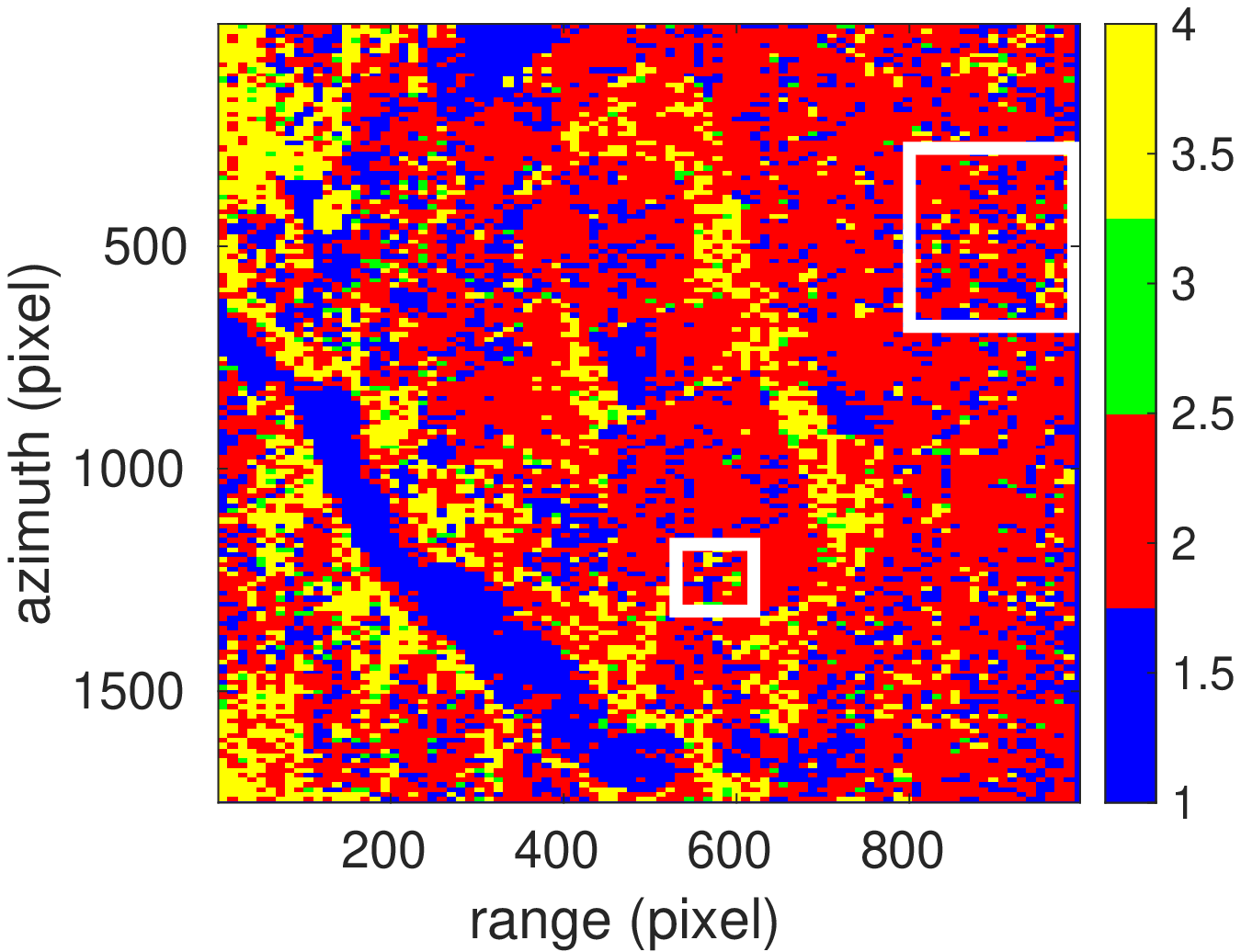}}
	\subfigure[BIC-D-P1]{\includegraphics[width=0.6\columnwidth]{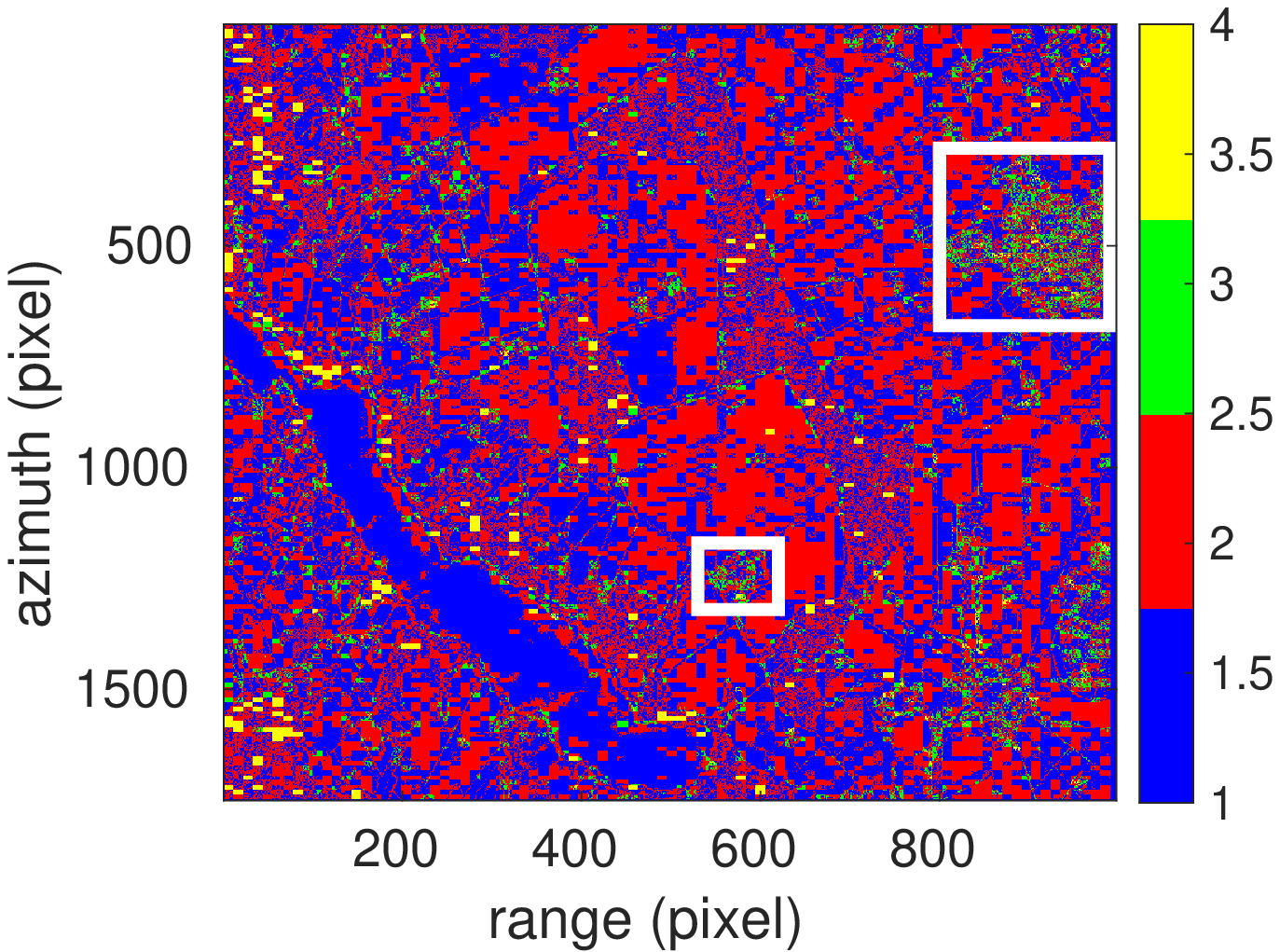}}
		\subfigure[BIC-D-P2]{\includegraphics[width=0.6\columnwidth]{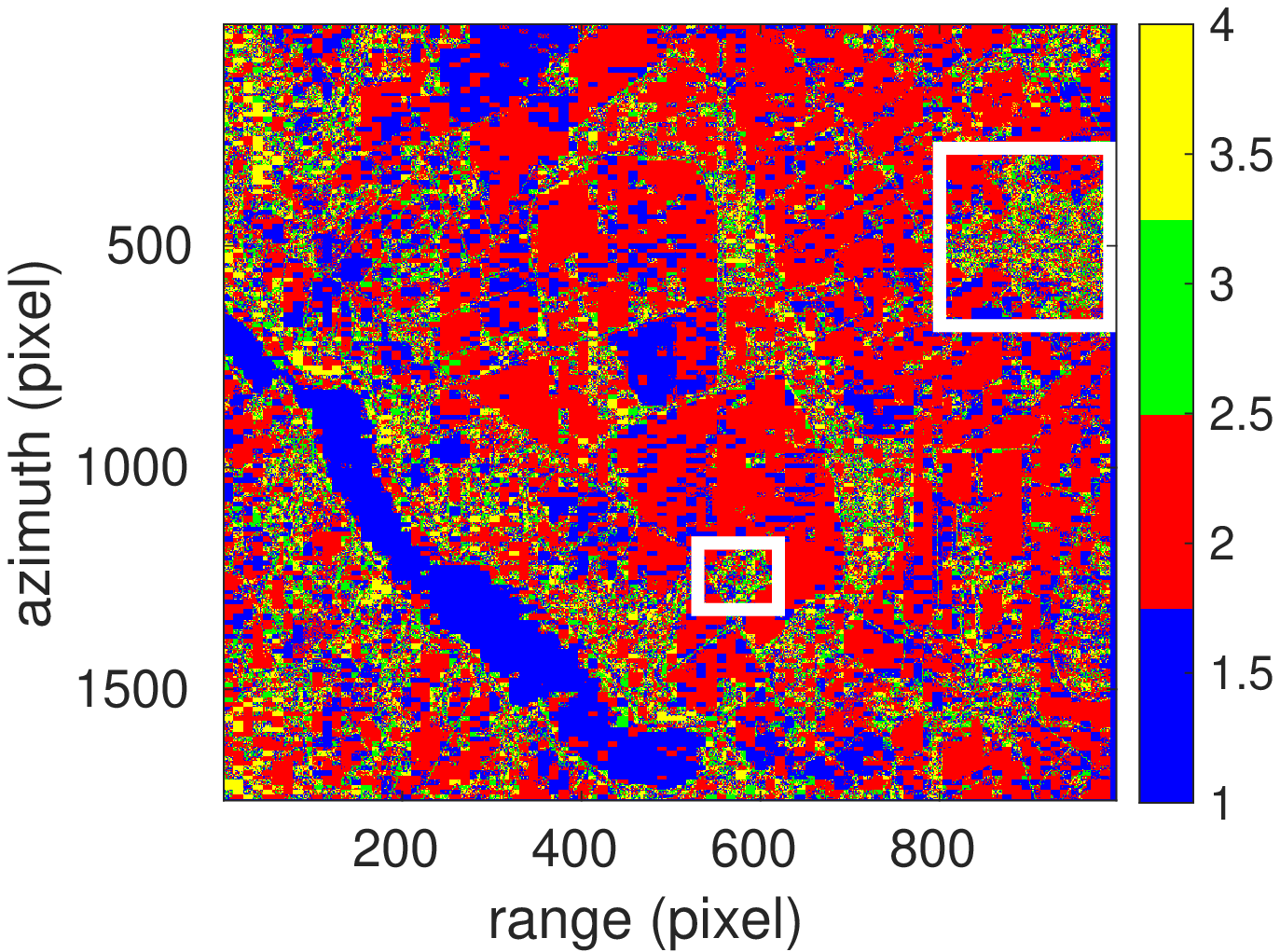}}
		\subfigure[GIC-based competitor]{\includegraphics[width=0.6\columnwidth]{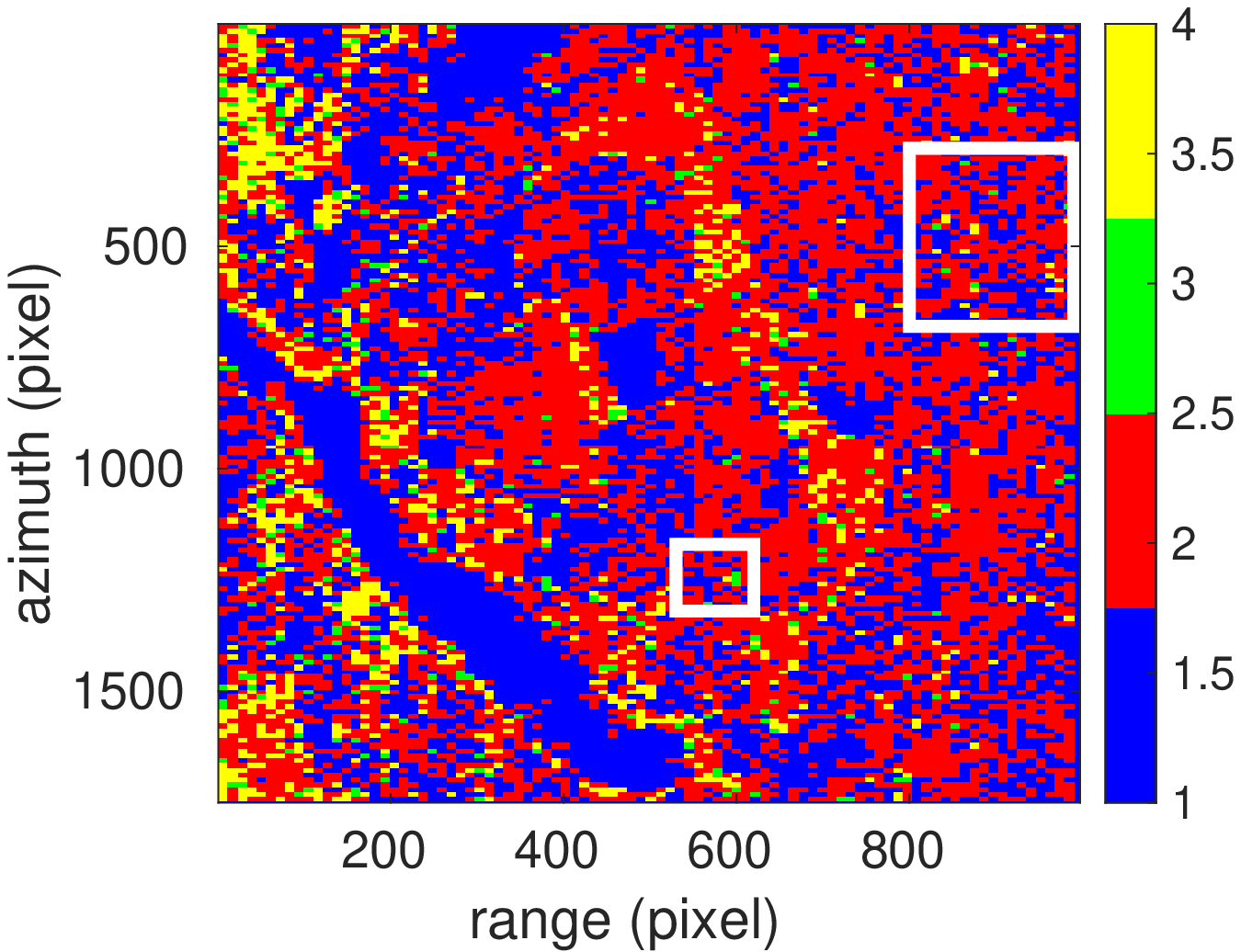}}
	\subfigure[GIC-D-P1]{\includegraphics[width=0.6\columnwidth]{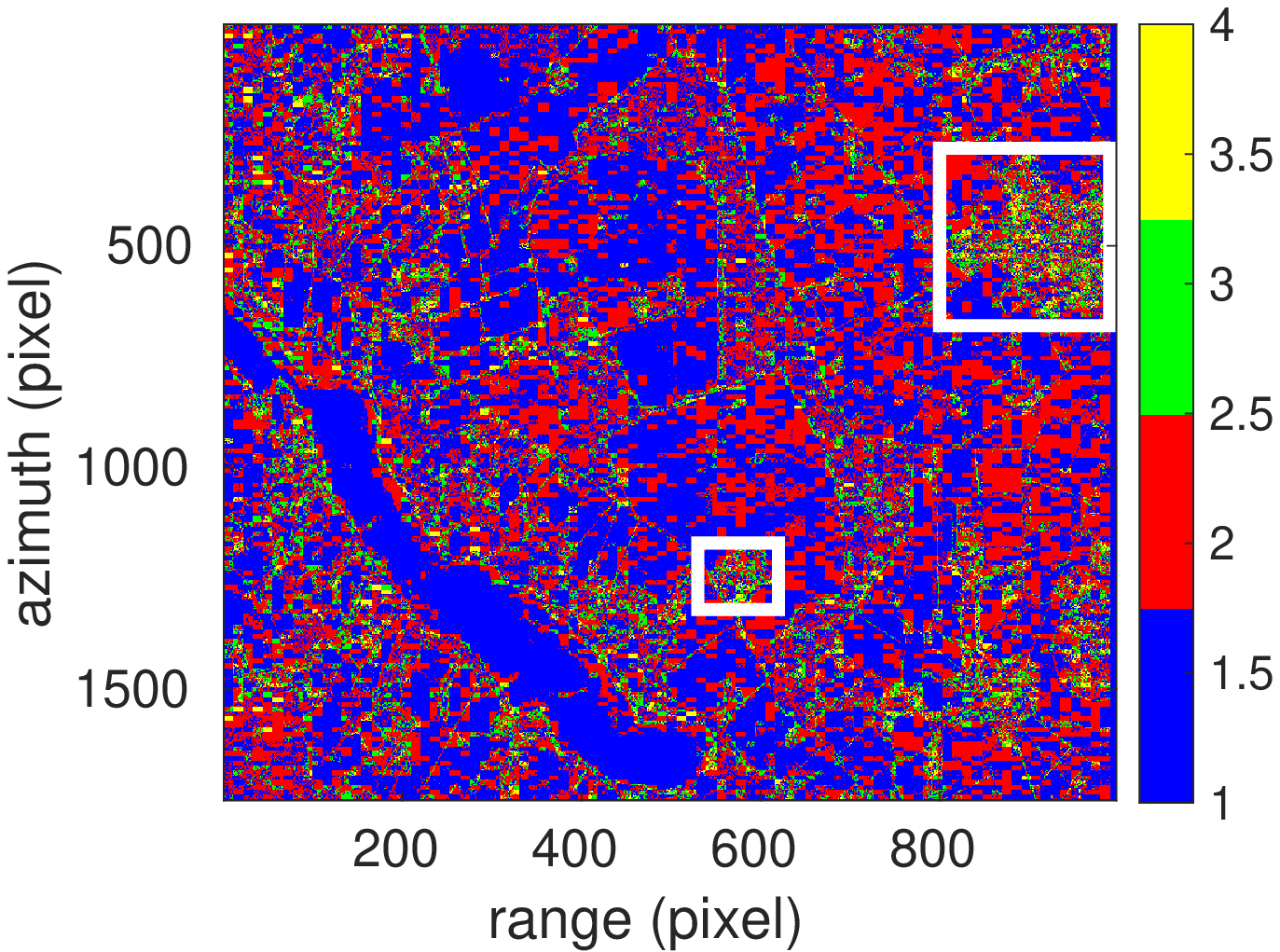}}
		\subfigure[GIC-D-P2]{\includegraphics[width=0.6\columnwidth]{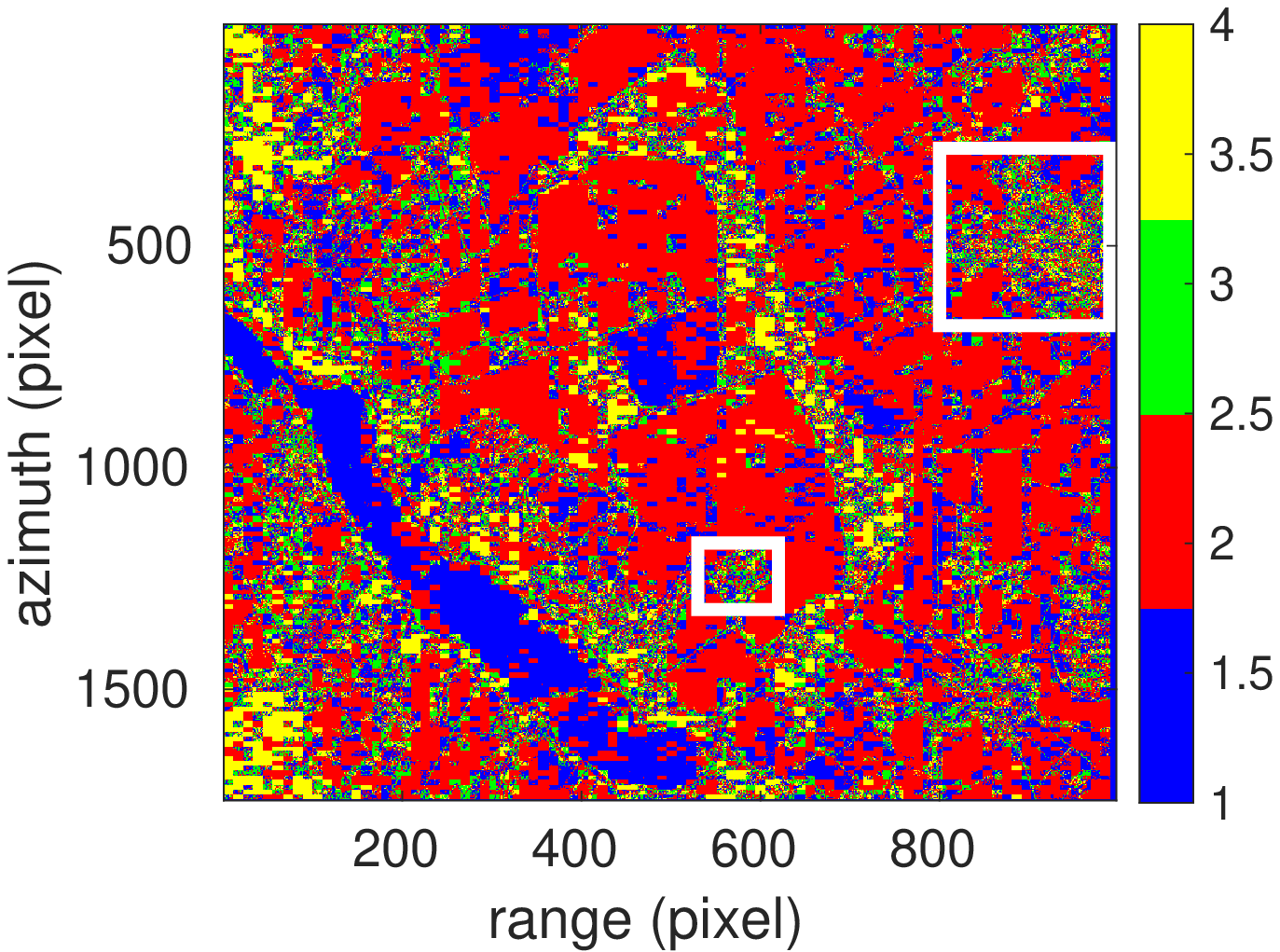}}
	\caption{Classification maps with real SAR data:  urban area of Tjele (small rectangle) and Orum (great rectangle).}
	\label{figure_real}
\end{figure*}

In Figure \ref{figure_real}, the classification maps for all the proposed architectures are shown along with the 
classification results of the competitor \cite{DetectionSymmetries}. A window of size $11 \times 11$ pixels 
is used\footnote{The window moves over the entire image without data overlapping between consecutive positions.} 
and the threshold is set to the value obtained with the synthetic simulations.
The figure clearly sheds light on the fact that the proposed architectures are capable of providing
enhanced details and finer resolutions with respect to the competitor due to the inherent best classification capabilities.
In all the considered cases, the absence of symmetry (blue pixels) is revealed over the water. Red pixels indicating a 
detected reflection symmetry, in place of crops and bare fields, are predominant for the BIC- and GIC-like architectures. 
Yellow pixels (azimuth symmetry) are classified in the presence of forest areas. 
Rotation symmetries (green pixels) are very few for the classification maps obtained by the competitor, whereas, 
they are more present in the results obtained with the proposed architectures and appear in the regions containing 
buildings (for example in the two highlighted urban areas) and roads (that are more clearly visible for the proposed architectures 
with respect to the competitor).

\section{Conclusions}
\label{sec:conclusions}
In this paper, we have addressed the problem of detecting and classifying PCM structure variations within a data window moving over
a polarimetric SAR image. Unlike existing classification procedures that assume a specific PCM structure for all
vectors belonging to the sliding window, in this case, data might exhibit different unknown PCM structures. More importantly, 
the partition of the entire data set according to the respective PCM structures is unknown and must be estimated. 
This problem naturally leads to a multiple hypothesis test with only one null hypothesis and multiple alternative hypotheses.
In order to avoid a significant computational load, we have devised a design framework, grounded on hidden random variables,
which assign a PCM label to data vectors, and the EM algorithm tailored to the considered PCM structures.

The performance analysis, conducted on both simulated and real-recorded data also in comparison with
a suitable competitor, has highlighted that AIC-D-P1, GIC-D-P1, and AIC-D-P2 are capable of providing an excellent 
compromise between detection and classification performance under all the considered hypotheses and for $K>120$. 
In addition, if we restrict the set of hypotheses of interest, other architectures can guarantee good classification/detection 
performance at least for values of $K$ greater than $120$.

Future research tracks might encompass the extension of such architectures to the heterogeneous environment where the reflectivity
coefficient within the window under investigation is not spatially stationary.
\bibliographystyle{IEEEtran}
\bibliography{group_bib}
\end{document}